\documentclass[English,titlepage, 12pt]{article}

\usepackage{epsf}

\usepackage{amsmath,bm}
\usepackage{amssymb}
\usepackage{graphicx}
\usepackage{amsfonts}
\usepackage{slashed}
\usepackage{color}

\usepackage{epsfig}
\usepackage{setspace}
\usepackage{ifthen}
\usepackage{graphicx}
\usepackage{amssymb}
\usepackage{xypic}
\usepackage{lscape}
\usepackage{amsthm}

\definecolor{red}{rgb}{1,0,0}

\newcommand{\be}{\begin{equation}}

\newcommand{\ee}{\end{equation}}
\newcommand{\bea}{\begin{eqnarray}}
\newcommand{\eea}{\end{eqnarray}}

\newcommand{\bra}{{\langle}}
\newcommand{\ket}{{\rangle}}
\newcommand{\tr}{\hbox{ Tr}}

 \newcommand{\myfig}[3]{\begin{figure}[ht]
\begin{center}
\leavevmode \epsfxsize=#2cm \epsfbox{#1}
\end{center}
\caption{#3} \label{fig:#1}
\end{figure}}

\begin{document}

\title{Gauge Invariance, Geometry and Arbitrage}
\author{Samuel E. V\'azquez \\
{\small Perimeter Institute for Theoretical Physics}, \\{\small  31 Caroline St. N., Waterloo, ON, Canada N2L 2Y5}\\
{\small Email: svazquez@perimeterinstitute.ca}  \\\\
Simone Farinelli, \\
{\small UBS Group Risk Methodology}
\\ {\small P.O. Box, Pelikanstrasse 6/8, CH-8098, Zurich}\\
   {\small Email: simone.farinelli@ubs.com}}

\maketitle

\begin{abstract}
In this work, we identify the most general measure of arbitrage for
any market model governed by It\^o processes. We show that our
arbitrage measure is invariant under changes of num\'{e}raire and
equivalent probability. Moreover, such measure has a geometrical interpretation  as
a gauge connection. The connection has zero curvature if and only if
there is no arbitrage. We  prove an extension of the Martingale
pricing theorem in the case of arbitrage. In our case, the present
value of any traded asset is given by the expectation of future
cash-flows discounted by a line integral of the gauge connection.  We
develop simple strategies to measure arbitrage
using both simulated and real market data. We find that, within our
limited data sample, the market is efficient at time horizons of one
day or longer. However, we provide strong evidence for non-zero arbitrage in
high frequency intraday data. Such events seem to have a decay time
of the order of one minute.

\end{abstract}

\section{Introduction}

The no-arbitrage principle is the cornerstone of modern financial
mathematics. Put it simply, an arbitrage opportunity allows an agent
to make a non-risky profit with zero or negative net investment (see
\cite{DeSc08}). Under the no-arbitrage assumption\footnote{In
continuous time the no-arbitrage condition (NA) has to be sharpened
to the no-free-lunch-with-vanishing-risk condition (NFLVR). Only
under the (NFLVR) the fundamental theorem of asset pricing can be
proved, see \cite{DeSc08}. Therefore, in this paper, when referring
to (NA) in continuous time, (NFLVR) is meant. } one can assign in a
complete market a unique price to the derivative of any traded
assets using the replicating portfolio method (see \cite{MuRu07} or
\cite{CvZa04}). The no-arbitrage principle can also be shown to be
equivalent to a weaker form of economic equilibrium (cf.
\cite{CvZa04}) and can therefore be seen as a form of market
efficiency (see \cite{Fam98}, \cite{Ma03}). It is then not
surprising that most financial and economic literature is based on
the no-arbitrage assumption.

Nevertheless, the no-arbitrage principle represents a very strong
assumption about market dynamics which must be tested empirically.
Even when the market participants use no-arbitrage models, the
ultimate price of any security which is traded in a centralized
market is set by supply/demand and the complex dynamics of the order
book. That is, ``the market" sets the prices. It then makes sense to
ask:  how efficient are these final prices? In order to answer this
question one needs a measure of arbitrage.

There is a large body of empirical studies on financial arbitrage.
Most of these studies focus on measuring the ``excess return" of
particular trading strategies (see e.g. \cite{JeTi93},
\cite{GaGoRo06}). Other studies try to find violations to general
no-arbitrage relations between option prices (see e.g.
\cite{AcTi99}). However, there does not seem to be a consensus on
whether the reported market ``anomalies" are due to arbitrage, or
simply to random fluctuations (see \cite{Ma03}, \cite{Fam98}). Part
of the problem is that there seems to be no general  measure of
arbitrage which can be applied to any traded asset.  One of the main
goals of this paper is to define such measure.

The second goal of this paper is to provide a geometrical
interpretation of the arbitrage measure. In particular, it has been
speculated long ago by Ilinski \cite{Il97}, \cite{Il01} and Young
\cite{Yo99} that arbitrage should be viewed as the ``curvature" of a
gauge connection, in analogy to some physical theories. The fact
that gauge theories are the natural language to describe economics
was first proposed by Malaney and Weinstein in the context of the
economic index problem  \cite{Ma96}, \cite{We06}. The need for such
mathematical language can easily be seen from the fact that prices
are only relational. More precisely, let $X(t) =
(X_1(t),X_2(t),\ldots)^{\dagger}$ be the price vector of all goods
in the economy at time $t$, in some common unit (say USD). Since the
measuring units are arbitrary, fundamental economic laws must be
invariant under the transformations \be\label{gauge} X(t)
\rightarrow \Lambda(t) X(t)\;,\ee where $\Lambda(t) > 0$ is a
positive stochastic factor\footnote{For example, $\Lambda(t)$ can be
the EUR/USD exchange rate.}. In physics, a (local) transformation
such as Eq. (\ref{gauge}) is known as a {\it gauge transformation}.
These represents a redundancy of our description of the economic
system. The laws of the economy should be gauge invariant. The need
for a gauge theoretical approach to economics was highlighted
recently in \cite{Sm09}. The role of gauge invariance in option
pricing has been studied in \cite{HoNe99a}, \cite{HoNe99b},
\cite{HoNe00} and \cite{HoNeVe01}. For an  unrelated use of
differential geometric methods in (no-arbitrage) option pricing see
\cite{La08}. A recent proposal for a gauge connection in finance,
and its relation to arbitrage, was presented  in \cite{Far08},
\cite{Far09a} and \cite{Far09b}. In fact, we will show that the
curvature of the gauge connection proposed in \cite{Far08},
\cite{Far09a} and \cite{Far09b} is equivalent to our arbitrage
measure.\par In physics, curvature is a {\it gauge invariant}
measure of the path dependency of some physical process. For
example, readers familiar with electrodynamics might recall the
vector potential $A_\mu$, where $\mu = 0,1,2,3$ label the space-time
directions. In differential geometry for theoretical physics,
$A_\mu$ is known as a {\it gauge connection}. Now consider a charged
particle which is traveling along some trajectory in space-time
$x^\mu(s)$, $s\in [0,1]$. The interaction of this particle with the
gauge potential is proportional to the line integral, $\int_\gamma A
:= \sum_\mu \int_0^1 A_\mu(x(s)) \dot x^\mu(s) ds$. Now suppose that
we make an infinitesimal change in the path of the particle $\delta
x^\mu(s)$, keeping the boundary conditions fixed: $\delta x^\mu(0) =
\delta x^\nu(1) = 0$. The interaction changes by $\delta \left(
\int_\gamma A \right) = \sum_{\mu,\nu}\int_0^1 \delta x^\mu(s) \dot
x^\nu(s) F_{\mu\nu}$, where $F_{\mu\nu} = \partial_\mu A_\nu -
\partial_\nu A_\mu$ is known as  the {\it curvature} of $A$.
Therefore, we see that for zero curvature $F_{\mu\nu} = 0$, the line
integral $\int_\gamma A$ is {\it independent of the path $\gamma$}.
Moreover, note that the curvature is invariant under a gauge
transformation of the form $A_\mu \rightarrow A_\mu + \partial_\mu
\Lambda$, where $\Lambda$ is any function of space-time.

We will find a very similar construction in the case of mathematical
finance.  In particular, we will show that the arbitrage curvature
defined in this paper measures the path dependency of the present
value of a self-financing portfolio of traded assets with fixed
final payoff. The no-arbitrage principle is then equivalent to a
zero-curvature condition.  In analogy with the electromagnetic
curvature $F$, we expect that any measure of arbitrage should be
invariant under the gauge transformation in Eq. (\ref{gauge}).
Moreover,  the {\it fundamental theorem of asset pricing} states
that the no-arbitrage principle is equivalent to the existence of a
probability measure with respect to which asset prices expressed in
terms of a num\'{e}raire are Martingales \cite{MuRu07}. Therefore,
we expect that any measure of arbitrage should also be invariant
under a change of  probability. These are, in fact, two very
important properties that will {\it characterize} our arbitrage
measure.

This paper takes a ``macroscopic" or phenomenological approach to
arbitrage. More precisely, we will study arbitrage from the point of
view of general stochastic models. We do not address the question of
what is causing the arbitrage. Our main goal is to identify the
gauge invariant financial observables that indicate an arbitrage
opportunity.  We believe that a proper understanding of such
quantity, including its geometrical interpretation,  is a first step
towards a theory of non-equilibrium economics. Moreover, our methods
can be applied to the construction of profitable trading strategies.
Our main assumption is that the prices of all financial instruments
can be described by It\^o processes.  Moreover, we ignore
transaction costs\footnote{Note that, as pointed out in
\cite{ShVi97}, many possible arbitrage opportunities disappear once
one takes into account market frictions such as transaction costs.
Therefore, it is important to keep in mind that  even when we
measure a  non-trivial curvature in the market, it does not mean
that it can always be exploited in a practical trading strategy.}.

The organization and main results of the paper are the following. In
section 2 we present the class of models that we use in the rest of
the paper. They are very general and include the case of stocks,
bonds and commodities, and more complicated derivative products. We
decompose the dynamics of these models in terms of their gauge
transformation properties with respect to Eq. (\ref{gauge}). We
identify the gauge invariants and show that they represent an
obstruction to the existence of a Martingale probability measure. We
conclude section 2 with an example with three assets, and we derive
a modified non-linear Black Scholes equation with arbitrage.

In section 3, we give a geometrical interpretation to the gauge
invariant quantities defined in section 2. Our main goal is to
identify the stochastic gauge invariants of section 2 with the
curvature of a gauge connection. We begin with a review of the
Malaney-Weinstein connection \cite{Ma96}, \cite{We06}, which is done
in the context of differentiable economic paths. In section 3.1, we
generalize the Malaney-Weinstein construction to stochastic
processes and prove an asset pricing theorem. The main result of
this section is that the present value of any self-financing
portfolio of traded assets is given by the conditional expectation
of future cash-flows, discounted by a line integral of the
Malaney-Weinstein gauge connection. We show how the value functions
of different portfolios replicating the same contingent claim are
related to the arbitrage curvature. Finally, we show that the gauge
connection recently proposed in \cite{Far08}, \cite{Far09a} and
\cite{Far09b}, is equivalent to the present construction. Readers
interested in only the arbitrage measure and the detection
techniques, can skip section 3. None of the results of section 4
require an understanding of the geometry of arbitrage. In section 4
we develop a simple algorithm to measure the arbitrage curvature
using financial data. We explain the main sources of error in such
measurement. The algorithm is applicable to any financial
instrument. We provide examples with financial data of stock indexes and index futures. We find that, on long time scales, the market is very
efficient. However, we provide strong evidence for non-zero
curvature fluctuations at short time scales in the order of one
minute. We conclude in section 5.

\section{Stochastic Models and Gauge Invariance}
Let ${\cal M}=\{0,1,2,\dots,N-1\}$ be the set of all traded securities at
any point in time in the market. We will use greek indices
$\mu,\nu,\ldots$ to label members of the set ${\cal M}$. We denote
the price of security $\mu \in {\cal M}$ by $X_\mu$.  Our main
assumption is that the dynamics of all such securities is described
by It\^o processes of the form \cite{So06} and \cite{Sh00}:
 \be \label{Ito} dX_\mu := X_\mu \left( \alpha_\mu dt + \sum_a \sigma^a_\mu  dW_a\right)\;,\;\;\; \forall \mu \in {\cal M} \;,\ee
 where $W_a$ are standard Brownian motion such that $W_a(t)  - W_a(0)$ are independently and normally distributed random variables with,
 \be \mathbb{E}\left[ W_a(t) - W_a(0)\right] = 0\;,\;\; \text{Cov}\left[W_a(t) - W_a(0), W_b(t) - W_b(0)\right]  = t \delta_{a b}\;.\ee

We make no assumptions about the number of Brownian terms, and hence
the completeness of the market. The set of Brownian motions
  $\{ W_a\}$ represent all the  randomness in
the market. Therefore, they induce a natural
filtration $\mathcal{F} = (\mathcal{F}_t)_{t\geq 0}$ for our probability
space. The coefficients $\alpha_\mu$ and $\sigma^a_\mu$ can also be
stochastic processes adapted\footnote{ In simple terms, a process $p$ adapted to ${\cal F}$ means that it does not depend on future values of the Brownian motion. In other words, $p(t)$ can only depend on $\{ W_a(s)\}$ up to time $s\leq t$.
} to the filtration $\mathcal{F}$.
 However, they are assumed to satisfy suitable conditions to ensure the
existence of the price processes $X_\mu$ (see \cite{LaLa06}).   This
class of models is very general and includes stocks, bonds, options,
etc. Moreover, the case of fat tails in the distribution of returns
is also included, since this is known to be generated by stochastic
volatilities $\sigma^a_\mu$.

Looking back at Eq. (\ref{Ito}), we can see that the tangent space $dX_\mu$ has a natural decomposition into the directions which contain all the randomness ($  \sum_a \sigma^a_\mu  dW_a$) and those orthogonal to it.
 Therefore, we will make the following decomposition of the drift term in Eq.
(\ref{Ito}): \be \label{alphamu} \boxed{\alpha_\mu = \alpha + \sum_a
\beta^a \hat\sigma_\mu^a  + \sum_{A \in \cal N}  \alpha^A J^A_\mu\;,
}\ee where $\cal N$ is the space spanned by basis vectors
$J^A:=[J^A_0,\dots,J^A_{N-1}]^{\dagger}$ such that \be \label{Jprop}
\sum_\mu J^A_\mu J^B_\mu = \delta^{A B}\;,\;\;\; \sum_\mu J^A_\mu =
0\;,\;\;\; \sum_\mu J^A_\mu \sigma_\mu^a = 0\;, \;\;\; \forall
a\;,\ee and
\begin{equation}
\alpha :=  \frac{1}{N}\sum_\mu \alpha_\mu, \quad \alpha^A  := \sum_\mu J^A_\mu
\alpha_\mu\quad
\hat\sigma^a_\mu:=\sigma^a_\mu-\sigma^a,\quad\sigma^a:=\frac{1}{N}\sum_{\mu}\sigma^a_\mu.
\end{equation}
We will refer to ${\cal N}$ as the {\it null space} of the market. Note that this space is orthogonal to all the randomness in the tangent space $dX_\mu$.
However, we need to remember that  ${\cal N} $ might be trivial. The definition
of $\beta^a$ in Eq. (\ref{alphamu}) is not unique if the vectors
$\hat \sigma^a_\mu:=[\hat \sigma^a_0,\dots,\hat
\sigma^a_{N-1}]^{\dagger}$ are linearly dependent. This is the case
of, for example, an incomplete market with more Brownian motions
than traded securities. Moreover, $\alpha^A$ is unique up to
rotations in the null space. As we will see, the quantities
$\alpha^A$ are the unique gauge invariant measures of arbitrage. The
two main goals of this paper are to give a geometric interpretation
to the parameters $\alpha^A$, and to set up a procedure to measure
them using financial data.

Since prices are relative and only reflect an exchange rate between
two products, the units used to measure $X_\mu$ are arbitrary.
Therefore, the dynamics of the market must be invariant under a
change of measuring units.  In mathematical finance, this is known
as a change of num\'eraire \cite{MuRu07}, and it can be interpreted as
a {\it  gauge transformation}, \be X_\mu(t) \rightarrow \Lambda(t)
X_\mu(t)\;,\ee where $\Lambda$ is a positive stochastic process
which is adapted to the filtration $\mathcal{F}$. Another symmetry,
which is special to the particular models of Eq. (\ref{Ito}), is a
transformation of the probability measure. This is not really a
gauge symmetry, but corresponds rather to a change of variables of
the form $W_a(t) \rightarrow W_a(t) + \int^t \delta\beta^a(s) ds$.
\par Our next task is to study the transformation properties of the
different terms in Eqs. (\ref{Ito}) and (\ref{alphamu}).  The
following result follows.

\noindent {\bf Proposition 2.1:} {\it Consider a change of num\'eraire
of the form $ X_\mu \rightarrow \Lambda X_\mu$, where $\Lambda$ is a
positive stochastic process adapted to the filtration $\mathcal{F}$,
and $d\Lambda := \Lambda( \delta \alpha dt + \sum_a \delta \sigma^a
dW_a)$. Then, the coefficients of the It\^{o} processes, Eqs.
(\ref{Ito}) and (\ref{alphamu}), transform as \be \label{xform1}
\alpha \rightarrow \alpha + \delta \alpha + \sum_a \sigma^a
\delta\sigma^a \;,\;\;\; \beta^a \rightarrow \beta^a + \delta
\sigma^a \;,\;\;\; \sigma^a \rightarrow \sigma^a + \delta
\sigma^a\;. \ee Finally, under a transformation of the probability
measure given by the Radon-Nykod\'{y}m derivative
${d\mathbb{P}}/{d\mathbb{P}^*} = \exp\left(-\int^t\sum_a\delta\beta^a
dW_a-\frac{1}{2}\int^t\sum_a(\delta\beta^a(s))^2 ds\right)$, we have
the mapping of standard Brownian motions
\begin{equation}
W_a(t) \rightarrow W_a(t) + \int^t \delta\beta^a(s) ds
\end{equation}
and
 \be \label{xform2} \alpha
\rightarrow \alpha + \sum_a \sigma^a \delta \beta^a\;,\;\;\; \beta^a
\rightarrow \beta^a + \delta \beta^a\;.\ee In particular, it follows
that $\hat \sigma^a_\mu$, $\alpha^A$ and $J^A_\mu$ are invariant
under such transformations.}
\smallskip

\noindent {\bf Proof:}  The result in Eq. (\ref{xform1}) above follow from a simple application of It\^o rule to the product $X'_\mu := \Lambda X_\mu$:
\bea dX'_\mu &=&  d\Lambda X_\mu + \Lambda dX_\mu + d\bra \Lambda,X_\mu\ket \nonumber \\
&=& X'_\mu \left[ \left( \alpha + \delta\alpha +  \sum_a \sigma^a \delta\sigma^a+ \sum_a (\beta^a + \delta \sigma^a) \hat \sigma^a_\mu + \sum_{A \in {\cal N}} \alpha^A J^A_\mu \right)dt  \right. \nonumber \\
&&\left. + \sum_a (\hat \sigma^a_\mu + \sigma^a+ \delta\sigma^a) dW_a \right]\;,
\eea
where $d\bra \Lambda, X_\mu\ket = dt \Lambda X_\mu  \sum_a \delta\sigma^a \sigma^a_\mu $ is the differential of the quadratic variation.
The transformation in Eq. (\ref{xform2}) follows from a simple differentiation of $W_a$ in Eq. (\ref{Ito}):
\bea dX_\mu &=& X_\mu\left[ \left(\alpha + \sum_a \sigma^a \delta \beta^a + \sum_a (\beta^a + \delta \beta^a) \hat \sigma^a_\mu + \sum_{A \in {\cal N}} \alpha^A J^A_\mu \right)dt  \right. \nonumber \\
&&\left. + \sum_a (\hat \sigma^a_\mu + \sigma^a) d W_a^* \right]\;,
\eea
where we defined $W_a(t) = W_a^*(t) + \int^t \delta\beta^a(s) ds$.
Note that both $\alpha^A$ and $J^A$ are unchanged by these gauge transformations. In particular, suppose that $\sum_\mu J^A_\mu \sigma^a_\mu = 0$. Then, it follows from Eq. (\ref{Jprop}) that $\sum_\mu J^A_\mu (\sigma^a_\mu  + \delta\sigma^a) = \sum_\mu J^A_\mu \sigma^a_\mu = 0$.

\noindent $ \square$

So far we have taken the existence of the basis vectors $J^A$ for granted.
A constructive procedure to find such basis, if non-trivial, is given by the following proposition.

\bigskip
\noindent {\bf Proposition 2.2:}\label{Prop22} {\it  Let $\Omega$ be
the symmetric and real $N \times N$ matrix with component
$\Omega_{\mu\nu} = \sum_a \sigma^a_\mu \sigma^a_\nu$, where $N =
\text{dim}({\cal M})$. Moreover, define $U$ as the matrix of all
ones, e.g. $U_{\mu\nu} = 1$, $\forall \mu,\nu \in \cal M$. Then, the
matrix $G$ defined by\be \label{G} G = \Omega  - \frac{1}{N} \left(
U \Omega + \Omega U\right) + \frac{1}{N^2} \tr(U \Omega) U\;,\ee is
gauge invariant. Let ${\cal N}_G$ be the null space of matrix $G$
such that $\sum_\mu J_\mu = 0$ for any non-trivial $J \in  {\cal
N}_G$. Then ${\cal N}_G =  {\cal N}$. In particular, the space
${\cal N}_G$ is spanned by the orthonormal zero-modes of $G$ which
are orthogonal to the vector $J = (1,1,\ldots,1)^{\dagger}$. }

\bigskip
\noindent {\bf Proof:} First we need to prove that the space of
vectors $J$ such that $ J^{\dagger}   \sigma^a = 0$, $\forall a$, is
in one-to-one correspondence with the zero modes of $\Omega$:
$\Omega  J = 0$.  Obviously, if $J^{\dagger}   \sigma^a = 0$, it
follows that $J$ is also a zero-mode of $\Omega$. To prove the
converse, suppose that $\Omega  J = 0$, but $J^{\dagger} \sigma^a  =
\lambda^a$, where $\lambda^a \neq 0$ for at least one value of $a$.
Then, $0=J^{\dagger}  \Omega   J  = \sum_a (\lambda^a)^2$, which can
only be true if $\lambda^a = 0$ $\forall a$.

Now we turn our attention to the matrix $G$, defined in Eq.
(\ref{G}). Using the gauge transformation $\sigma^a_\mu \rightarrow
\sigma^a_\mu + \delta \sigma^a$, we can see that $\Omega$ transforms
as \be \label{Omegaxform} \Omega_{\mu\nu} \rightarrow
\Omega_{\mu\nu} + \sum_a \delta\sigma^a(\sigma^a_\mu + \sigma^a_\nu)
+ \sum_a (\delta\sigma^a)^2\;.\ee It is then straightforward to
verify the gauge invariance of the matrix $G$. Next we recall that
the space ${\cal N}$ is spanned by (non-trivial) orthonormal
zero-modes of $\Omega$ such that they also satisfy $\sum_\mu J_\mu =
0$. One can define a similar space ${\cal N}_G$ for $G$. It is easy
to verify that any vector $J \in {\cal N}$ is also a vector in
${\cal N}_G$. On the other hand, for any vector $ J' \in {\cal
N}_G$, it follows from Eq. (\ref{G}) that $ \Omega   J' = 0$. Thus,
we have proven that ${\cal N} = {\cal N}_G$.

It is easy to verify that $\sum_\mu G_{\mu\nu} = 0$. Therefore, the
vector $J = (1,1,\ldots,1)^{\dagger}$ is a particular zero-mode of
G. Now take any other zero-mode of $G$, call it $J'$, which is
orthogonal to $J$. It follows that, $0 = J^{\dagger}  J' = \sum_\mu
J'_\mu$. Therefore, $J' \in {\cal N}_G = {\cal N}$. This completes
the proof.

\noindent $\square$

So far we have talked about the full set of securities of the market.
However, it is clear that the decomposition in Eq. (\ref{alphamu})
can be done for any subset of the market. That is, suppose we observe
a subset of the prices $X_i$, $i \in {\cal S} \subset \cal M$.
Moreover, suppose we find that within this subset one still can
find some zero-modes $J^A$ obeying, $\sum_i J^A_i \sigma^a_i = 0$,
$\forall a$ and $\sum_i J^A_i = 0$. One can then easily lift these vectors
to the full set ${\cal M}$ by taking $J^A_{\cal M} = (J^A, \vec 0)$. This
represents a particular choice of basis in the null space ${\cal N}$.
By observing a sub-sector of the market, we will only have access to some
of the components of $\alpha^A$. For notational convenience, we will not
distinguish between the full market and a subset of it in what follows.

Under the no-arbitrage assumption (see \cite{MuRu07},
\cite{CvZa04}), it is always possible to find a common positive
discount factor $\Lambda$ and an equivalent probability measure
$\mathbb{P}^* \sim \mathbb{P}$ such that the discounted prices
$\Lambda X_\mu$ are martingales: $\Lambda(t) X_\mu(t)  =
\mathbb{E}^*_t[\Lambda(T) X_\mu(T)]$, where $t \leq T$. This is known as the {\it Martingale representation theorem}
(see \cite{So06} and \cite{Sh00}).
In our
language, this means that there is a gauge transformation mapping
$X_\mu$ to $\mathbb{P}^*$-Martingales. In other words, if there is
no arbitrage, price processes are gauge-equivalent to
$\mathbb{P}^*$-Martingales for some probability measure
$\mathbb{P}^*$. The result of the Martingale representation theorem can only be obtained  if one is able to write $\int_t^T d(\Lambda X_\mu) := \int_t^T \gamma_\mu^a dW^*_a$ for some adapted process $\gamma^a_\mu$. The reason is that  the stochastic integral $\int_t^T \gamma_\mu^a dW^*_a$ is a Martingale: $\mathbb{E}_t^*\left[\int_t^T \gamma_\mu^a dW^*_a\right] = 0$.
By proposition 2.2 there is neither a change of
probability nor a choice of a positive discount factor for which the
vector  $\sum_{A \in {\cal N}} \alpha^A J^A$ is mapped to $0$ (in
contrast to $\alpha$ and all $\beta^a$s which can indeed be made
vanish). Therefore, it is easy to see that the term $\sum_{A \in {\cal N}}
\alpha^A J^A$ parametrizes the obstruction to the existence of a
Martingale probability measure for any discounted price process $\Lambda X_\mu$.

  As $\alpha^A$s are gauge invariant
quantities, one expects that they should be observables. In the next
section we will relate this quantity to a gauge connection and its
curvature. In section 4 we will show that such quantity can indeed
be observed, and we explain simple strategies to
measure it. Before concluding this section, it is instructive to a
study particular example with three assets.

\subsection{An Example}
Consider the case of three assets $X_\mu$, $\mu = 0,1,2$, where $X_0$ is a savings account and $X_1,X_2$ are some other risky assets. All prices are measured in the same common units. We will assume only one Brownian motion. Therefore, the dynamics of the prices is described by
\bea dX_0 &=& r X_0 dt\;,\nonumber \\
\label{example} dX_i &=& X_i \left[ \alpha_i dt + \sigma_i dW\right]\;,\;\;\; i = 1,2\;.\eea
For later convenience, we assume that the interest rate $r$ is deterministic.
In order to do the decomposition in Eq. (\ref{alphamu}) we need to find a basis for the null space ${\cal N}$. In this case, since there is only one Brownian motion and two risky assets, there will be only one null direction. To calculate it, we start by identifying the $\Omega$ matrix:
\be \Omega = \begin{pmatrix}
0 & 0 &0 \\
0& \sigma_1^2  & \sigma_1 \sigma_2 \\
0& \sigma_1 \sigma_2 & \sigma_2^2
\end{pmatrix}\;.\ee
One can now construct the $G$ matrix using Eq. (\ref{G}). The
explicit form of $G$ is not very illuminating. The unormalized
eigenvectors of $G$ are found to be \be V_1 = \begin{pmatrix}
2 -\frac{3 \sigma_1}{\sigma_1 + \sigma_2} \\
0\\
0
\end{pmatrix}\;,\;\;\; V_2 = \begin{pmatrix}
-1 +\frac{3 \sigma_1}{\sigma_1 + \sigma_2} \\
1\\
0
\end{pmatrix}\;,\;\;\; V_3 = \begin{pmatrix}
\frac{\sigma_1 + \sigma_2}{\sigma_1 - 2 \sigma_2}\\
\frac{\sigma_2 - 2 \sigma_1}{\sigma_1 - 2 \sigma_2}\\
1
\end{pmatrix}\;,
\ee where \be G  V_1 = 0\;,\;\;\; G  V_2 = 0\;,\;\;\; G  V_3 =
\frac{2}{3} (\sigma_1^2 + \sigma_2^2 - \sigma_1 \sigma_2) V_3\;. \ee
In order to find a basis for the null space $\cal N$ defined in Eq.
(\ref{Jprop}), we need to project $V_1$ or $V_2$ into the space
orthogonal to the vector $J^0 = (1,1,\ldots 1)^{\dagger}$. To do
this, we define the projection matrix \be P_U := \frac{1}{3} U
\;,\;\;\; P_U^2 = P_U\;,\ee where $U$ is the $3\times 3$ all-ones
matrix. Note that $\mathbf{1}- P_U$ projects into the space
orthogonal to $J^0$.  Our choice for the normalized null vector is
then \be \label{Jex} J = \frac{ (\mathbf{1} - P_U)  V_1}{\sqrt{ [(\mathbf{1} -
P_U)  V_1]^{\dagger}  (\mathbf{1} - P_U)  V_1}} = \frac{1}{\sqrt{2}
\sqrt{\sigma_1^2 + \sigma_2^2 - \sigma_1 \sigma_2 }}
\begin{pmatrix}
\sigma_1 - \sigma_2 \\
\sigma_2 \\
-\sigma_1
\end{pmatrix}\;.  \ee
It is easy to verify that $J$ obeys the properties given in Eq. (\ref{Jprop}).

We can now go back to the decomposition given in Eq. (\ref{alphamu}). Using Eqs. (\ref{example}), we find
\be \label{alpha0} \alpha  = r  - \beta \hat \sigma_0 - \tilde \alpha J_0\;,\;\;\; \sigma_0 = 0\;,\ee
where $\hat \sigma_\mu  = \sigma_\mu - \frac{1}{3} \sum_{\nu = 0}^2 \sigma_\nu$, and $\tilde \alpha$ is the arbitrage vector $\alpha^A$, which in this case has only one component $\alpha^1 := \tilde \alpha$.  Therefore, inserting Eqs. (\ref{alpha0}) into Eqs. (\ref{example}), we can write the evolution equations as
\bea \label{X0} dX_0 &=& r X_0 dt\;, \\
\label{X1} dX_1 &=& X_1 \left[ \left(r + \beta \sigma_1 + \tilde \alpha \frac{2 \sigma_2 - \sigma_1}{\sqrt{2}\sqrt{\sigma_1^2 + \sigma_2^2 -\sigma_1\sigma_2}}\right)dt + \sigma_1 dW\right]\;, \\
\label{X2} dX_2 &=& X_2 \left[ \left(r + \beta \sigma_2 + \tilde
\alpha \frac{\sigma_2 - 2 \sigma_1}{\sqrt{2}\sqrt{\sigma_1^2 +
\sigma_2^2 -\sigma_1\sigma_2}}\right)dt + \sigma_2 dW\right]\;.\eea
For $\tilde \alpha = 0$, Eqs. (\ref{X0}) - (\ref{X2}) reduce to the
familiar no-arbitrage Black-Scholes dynamics. As usual, $\beta$ is
interpreted as the market price of risk.  Note that in this example,
both risky assets are exposed to the same market risk factor $W$.
The {\it volatility} $\sigma_i$ measures the coupling to such risk.
Under the no-arbitrage assumption, both assets should give the same
expected return per unit of risk. This is $\beta$. However, we see
that if $\tilde \alpha\neq 0$, $X_1$ and $X_2$ have different
expected returns, even when they are exposed to the same risk. This
discloses an arbitrage opportunity.

There is a very interesting consequence of Eqs. (\ref{X0}) - (\ref{X2}) when $X_2$ is any function of $X_1$, e.g. an option. For simplicity, consider the case where the only time dependence in $\tilde \alpha$ is of the form $\tilde \alpha = \tilde \alpha(t,X_1)$, where $\tilde \alpha(t,X_1)$ is a differentiable function of $t$ and $X_1$. Moreover, the interest rate $r$ is assumed to be deterministic.   In this case one can derive a non-linear version of the Black Scholes equation with arbitrage. For ease of notation let $X_1:= X$. Under our assumptions we will have that $X_2 = V(t, X)$. Then, using It\^o rule we find
\bea dV &=& \partial_t V dt + \partial_X V dX + \frac{1}{2} \partial_X^2 V d\bra X \ket\nonumber \\
&=& V\left( \alpha_2 dt + \sigma_2 dW\right)\;,\eea where we
identify \be \label{alpha2} \alpha_2 = \frac{\partial_tV}{V} +
\alpha_1 X \frac{\partial_X V}{V} + \frac{1}{2} \sigma_1^2 X^2
\frac{\partial_X^2 V}{V}\;,\;\;\; \sigma_2 = \sigma_1 X
\frac{\partial_X V}{V} \;.\ee Comparing Eq. (\ref{alpha2}) with Eq.
(\ref{X2}) we find \be \label{alpha2new} \alpha_2 =  r + \beta
\sigma_2 + \tilde \alpha \frac{\sigma_2 - 2
\sigma_1}{\sqrt{2}\sqrt{\sigma_1^2 + \sigma_2^2 -\sigma_1\sigma_2}}
= \frac{\partial_tV}{V} + \alpha_1 X \frac{\partial_X V}{V} +
\frac{1}{2} \sigma_1^2 X^2 \frac{\partial_X^2 V}{V}\;,\ee where from
Eq. (\ref{X1}) \be \alpha_1 = r + \beta \sigma_1 + \tilde \alpha
\frac{2 \sigma_2 - \sigma_1}{\sqrt{2}\sqrt{\sigma_1^2 + \sigma_2^2
-\sigma_1\sigma_2}}\;.\ee Therefore, after some algebra Eq.
(\ref{alpha2new}) becomes a modified non-linear Black Scholes
partial differential equation: \be\label{BS}{ \boxed{\partial_t V +
r X
\partial_{X} V + \frac{1}{2} \sigma^2_1 X^2 \partial_X^2 V
+\left(\sqrt{2} \tilde \alpha  \left[ 1 + \frac{ X \partial_X
V}{V}\left( \frac{X
\partial_X V}{V} - 1\right) \right]^{1/2}  - r\right) V= 0\;.}}\ee
Note that for $\tilde \alpha = 0$ this reduces to the familiar Black
Scholes equation. The non-linear Black Scholes equation is a special
case of the more general pricing theorem presented in section 3.

It is important to remember that the arbitrage parameter, $\tilde
\alpha$ in Eq. (\ref{BS}), can in general depend on time and the
stock price. Therefore, in principle almost any deformation of the
option price is possible.  It follows that the Eq. (\ref{BS}) can be
solved only if the arbitrage dynamics is known.   For example, consider the case where we set
\be \tilde \alpha := \frac{1}{2^{3/2} }(\tilde \sigma_1^2 - \sigma_1^2) \frac{X^2 \partial_X^2 V}{V \left[ 1 + \frac{ X \partial_X
V}{V}\left( \frac{X
\partial_X V}{V} - 1\right) \right]^{1/2}}\;,\ee
for some constant $\tilde \sigma_1^2$. Then, the option price obeys the usual Black-Scholes equation but with the ``wrong" volatility:
\be \partial_t V +
r \left(X
\partial_{X} V - V\right) + \frac{1}{2} \tilde \sigma^2_1 X^2 \partial_X^2 V = 0\;.\ee
This is a simple example of the well-known {\it volatility arbitrage}.

\section{The Gauge Connection}
The application of differential geometric ideas in economics can be
traced to the work of Malaney and Weinstein (\cite{Ma96},
\cite{We06}). It was found that the solution to the apparent
discrepancy among different economic growth indices could be solved
by the appropriate choice of a {\it covariant derivative}.  Such
derivative has the property that a self-financing basket of goods is
seen as ``constant". More technically, a self-financing basket is
interpreted as being ``parallel transported"  along a one
dimensional curve in the a base manifold spanned by prices and
portfolio nominals.   Then, there is a natural geometric index to
measure the growth of such basket, which was shown to be identical
to the so-called Divisa Index.
It is very illuminating to review this construction to gain
intuition about the relation between arbitrage and curvature. In
what follows, all quantities are assumed to be deterministic and
differentiable.  We will return to the stochastic case in the next
subsection.

A covariant derivative induces a connection one-form in
the base space, and in \cite{Ma96}, \cite{We06} this connection is
given by \be \label{A} A = \frac{\sum_\mu \phi_\mu dX_\mu}{\sum_\nu
\phi_\nu X_\nu}\;,\ee where $\phi_\mu$ are the portofolio nominals,
$V = \sum_\mu \phi_\mu X_\mu$, and the base space is parametrized by
the coordinates $(t,\phi_\mu,X_\mu)$.  Note that under a change of
num\'{e}raire $X_\mu \rightarrow \Lambda(X) X_\mu$, the connection
transforms as \be A \rightarrow A + d\Lambda\;.\ee  This is the
analog of the transformation rule of the vector potential in
electrodynamics.

A self-financing portfolio can be seen as being parallel transported
with the connection $A$ as \be \nabla_{\dot{\gamma}} V = \left. (d -
A) V \right|_{\dot{\gamma}} = 0\;,\ee where $\nabla_{\dot{\gamma}}$
is the covariant derivative along the trajectory $\gamma$. The
solution to this equation is simply, \be \label{Divisa}
\frac{V(T)}{V(t)}  = e^{\int_\gamma A}:= D_\gamma \;,\ee where
$\gamma$ is a particular self-financing trajectory
$(s,\phi(s),X(s))$ , $s\in [t,T]$, and $D_\gamma$ is known as the
Divisa Index.

The dependence of $D_\gamma$ on the choice of curve $\gamma$ is
parametrized by the {\it curvature} of the gauge connection, which
is given by \be\label{R} R = dA = \frac{1}{\left(\sum_\mu \phi_\mu
X_\mu\right)^2} \sum_{\nu,\sigma} \left( \phi_\nu X_\nu d\phi_\sigma
\wedge dX_\sigma - \phi_\sigma X_\nu d\phi_\nu\wedge
dX_\sigma\right)\;. \ee Note that the curvature is invariant under a
gauge transformation, as $d(A + d\Lambda) = dA$. In the
approximation where economic agents are price takers, the price
trajectory $X(t)$ is given exogenously, and we are only allowed to
make changes in the portfolio nominals $\phi$.  In other words, we can  write $dX_\mu = \dot X_\mu dt$ in Eq. (\ref{R}). One can then
restrict the curvature to the submanifold corresponding to the
$(t,\phi_{\mu})$ coordinates. The induced curvature in this
submanifold is given by \be R = \frac{1}{\left(\sum_\mu \phi_\mu
X_\mu\right)^2} \sum_{\nu, \sigma} \phi_\sigma X_\nu X_\sigma
\left(\frac{\dot X_\nu}{X_\nu} - \frac{\dot
X_\sigma}{X_\sigma}\right) d\phi_\nu \wedge dt := \sum_\mu
R_{\mu,t}d\phi_\mu\wedge dt\;.\ee In this case, the path dependency
of the Divisa Index, Eq. (\ref{Divisa}), can be written as, \be
\label{DivisaGamma}\delta_\gamma \log D_\gamma =  \sum_\mu \int_t^T
ds R_{\mu,t}(s) \delta\phi_\mu(s)\;,\ee where $\delta_\gamma$
represents a variation to the trajectory of the portfolio nominals.
Therefore, we see that Eq. (\ref{Divisa}) is independent on the path
$\gamma$ only if the price trajectories obey the zero-curvature
condition $R_{\mu,t} = 0 \implies \dot X_\mu(t) = \alpha(t)
X_\mu(t)$, $\forall \mu$.  The zero-curvature condition implies that
the prices of all securities evolve by the same common inflation
factor.

The relation between curvature and arbitrage goes as follow. Suppose
that the prices obeyed the zero-curvature condition given above. It
follows that, for any self-financing portfolio, we have \be V(T) =
V(t) e^{\int_\gamma A} = V(t) e^{\int_t^T\alpha(s)\, ds}\;,\ee for
$T \geq t$. In particular, if $V(t) = 0$ it follows that $V(T) = 0$.
Therefore, it is not possible to make wealth without a positive
initial investment. On the other hand, suppose that the curvature is
not zero. Consider two portfolio trajectories $\gamma_1$ and
$\gamma_2$ such that, say $D_{\gamma_1} > D_{\gamma_2}$ at some time
$T \geq t$, for the same initial wealth
$V_{\gamma_1}(t)=V_{\gamma_2}(t) > 0$. Now construct the difference
portfolio with nominals $\phi:=\phi_1-\phi_2$  and wealth
function\be V = V_{\gamma_1} - V_{\gamma_2}\;.\ee Then, at time $T
\geq t$ we have \be V(T) = \left(D_{\gamma_1} - D_{\gamma_2}\right)
V_{\gamma_1}(t) > 0\;,\ee while $V(t)=0$. In other words, we have
made wealth out of nothing.  In the next section we show how this
construction carries over to the stochastic case.

\subsection{The Stochastic Gauge Connection}
In the previous section we illustrated the relation between
curvature, path dependency and arbitrage, using the
Malaney-Weinstein connection. However, this construction only works
for differentiable economic trajectories in the base space $(
\phi,X)$. Nevertheless, we have found a direct analog of the
Malaney-Weinstein connection for It\^o processes, which we summarize
in the following theorem. In order to avoid technical complications,
we restrict our attention to an economy on a finite interval of time
$t \in [0,T]$.

\bigskip
\noindent {\bf Theorem 3.1}: {\it Consider any self-financing
portfolio $V = \sum_\mu \phi_\mu X_\mu$, so that $dV = \sum_\mu
\phi_\mu dX_\mu$. Then, there exist a (non-unique) equivalent
probability measure $\mathbb{P}^* \sim \mathbb{P}$  under which the price processes obey
\be \label{Xstar} dX_\mu = X_\mu \left[ \left(\alpha^*  + \sum_A \alpha^A J^A_\mu\right)dt + \sum_a \sigma^a_\mu dW_a^*\right]\;.\ee
Moreover, the
present value of $V(t)$ given some final payoff $V(T)$, $T\geq t$,
is given by \be \label{th1} V_\gamma(t)  = \mathbb{E}_t^*\left[ V(T)
e^{-\int_\gamma \Gamma}\right]\;,\ee {\it where $\gamma$ is some
self-financing trajectory,  and  $\Gamma$  is given by the
expectation of the Malaney-Weinstein connection, } \be \label{gamma}
{\Gamma =   \mathbb{E}_t^* \left[  \frac{\sum_\mu \phi_\mu
dX_\mu}{\sum_\nu \phi_\nu X_\nu}\right] =  \frac{\sum_{\mu,
A\in{\cal N}}\alpha^A J^A_\mu \phi_\mu X_\mu }{ \sum_\nu \phi_\nu
X_\nu} dt  + \alpha^* dt\;.} \ee Finally, the path dependency of the
present value of the portfolio, with fixed final payoff,  is
parametrized by} \be \label{th2} \delta V_\gamma(t) = - \sum_\mu
\int_t^T ds \mathbb{E}_t^*\left[ V(T) e^{-\int_\gamma \Gamma} \delta
\phi_\mu(s) R_{\mu,t}(s) \right]\;,\ee {\it where $R_{\mu,t}$ are
the components of the curvature two-form defined in the reduced base
space $(t,\phi)$, \be R = d\Gamma =  \frac{1}{\left(\sum_\mu
\phi_\mu X_\mu\right)^2} \sum_{\nu,\sigma,A \in {\cal N}}  \alpha^A
X_\nu X_\sigma \phi_\sigma \left(J^A_\nu - J^A_\sigma\right)
d\phi_\nu \wedge dt := \sum_\mu R_{\mu, t} d\phi_\mu \wedge
dt\;.\ee}

\noindent {\bf Proof:}  We start by writing the portfolio return as
\be dV = \sum_\mu \phi_\mu dX_\mu:= V\left( a dt + \sum_a b^a dW_a
\right) \;,\ee where \be a = \frac{\sum_\mu \alpha_\mu \phi_\mu
X_\mu}{\sum_\nu \phi_\nu X_\nu}\;,\;\;\;  b^a =  \frac{\sum_\mu
\sigma^a_\mu \phi_\mu X_\mu}{\sum_\nu \phi_\nu X_\nu}\;.\ee Now
consider the combination $V' := \Lambda V$, where we take (c.f. Eq.
(\ref{Ito}))
\be d\Lambda = \Lambda\left[\left(- a + \sum_a b^a
\beta^a\right)dt - \sum_a \beta^a dW_a  \right]\;.\ee A simple
application of It\^o rule gives \be dV' = V' \sum_a (b^a - \beta^a)
dW_a\;. \ee It is well known that any stochastic integral of the
form $\int_0^t \gamma dW_a$ is a Martingale \cite{So06} and
\cite{Sh00}. Therefore, we have \be \label{Vexp} V(t)  =
\mathbb{E}_t\left[ V(T) e^{\int_t^T d\log\Lambda }\right]\;.\ee A
further application of It\^o rule gives
 \be \label{dlog} d\log
\Lambda =  - \Gamma - \frac{1}{2} \sum_a (\beta^a)^2 dt - \sum_a
\beta^a dW_a\;,\ee
where $\Gamma$ is defined in Eq. (\ref{gamma}), with $\alpha^* := \alpha - \sum_a \beta^a \sigma^a$.

Now consider making a change of probability measure such that $W_a := W_a^*  - \int^t \beta^a(s) ds$.  It is easy to see that, under $\mathbb{P}^*$,  the price processes will obey Eq. (\ref{Xstar}) of the theorem. Moreover, the Radon-Nykod\'ym derivative is given by
\be \label{dPdPs} \frac{d\mathbb{P}}{d\mathbb{P}^*} = \exp \left[ -\frac{1}{2} \sum_a \int_t^T (\beta^a(s))^2 ds + \sum_a \int_t^T \beta^a dW_a^* \right] \;.\ee
Therefore, using Eqs. (\ref{dlog}) and (\ref{dPdPs}) in (\ref{Vexp}) we get
\bea V(t)  &=&
\mathbb{E}_t\left[ V(T) e^{- \int_\gamma \Gamma - \frac{1}{2} \sum_a \int_t^T (\beta^a)^2 dt - \sum_a\int_t^T
\beta^a dW_a }\right]\nonumber \\
&=&
\mathbb{E}_t^*\left[ V(T)  \frac{d\mathbb{P}}{d\mathbb{P}^*}  e^{- \int_\gamma \Gamma + \frac{1}{2} \sum_a \int_t^T (\beta^a)^2 dt - \sum_a\int_t^T
\beta^a dW_a^* }\right]\nonumber \\
&=& \mathbb{E}_t^*\left[ V(T)  e^{- \int_\gamma \Gamma  }\right]\eea

In order to prove that $\Gamma$ can be written as an expectation of
the Malaney-Weinstein connection, we recall that
\bea \Gamma  &=&  \sum_\mu  \frac{\phi_\mu(t) X_\mu(t)}{\sum_\nu \phi_\nu(t) X_\nu(t)}  \left(\alpha^* + \sum_{A \in \cal N} \alpha^A J^A_\mu \right)dt \nonumber \\
&=& \lim_{\delta t \rightarrow 0} \sum_\mu  \frac{\phi_\mu(t)}{\sum_\nu \phi_\nu(t) X_\nu(t)}  \left(\frac{\mathbb{E}_t^*[X_\mu(t+\delta t)] - X_\mu(t)}{\delta t}\right) dt \nonumber \\
&=& \lim_{\delta t \rightarrow 0}  \mathbb{E}_t^* \left[  \sum_\mu\frac{ \phi_\mu(t)}{\sum_\nu \phi_\nu(t) X_\nu(t)}\left(\frac{ X_\mu(t+\delta t) - X_\mu(t)}{\delta t} \right) dt \right] \nonumber \\
&=&  \mathbb{E}_t^* \left[  \frac{\sum_\mu \phi_\mu dX_\mu}{\sum_\nu
\phi_\nu X_\nu}\right] \;. \eea The last result of the theorem, Eq.
(\ref{th2}), follows simply by making a small change in the
portfolio nominals, and keeping the boundary conditions on $V$
fixed.

\noindent $\square$

Note that  the curvature of $\Gamma$ is zero if and only if
$\alpha^A = 0$, which is equivalent to the no-arbitrage condition.
Moreover, the probability measure $\mathbb{P}^*$ might not be
unique, as the choice of $\beta^a$ in general is not. This also implies that $\alpha^*$ is not unique in general.

\bigskip

\noindent A special case of a self-financing portfolio is a
portfolio containing just one base asset.\\
\noindent {\bf Corollary 3.2}: {\it For all assets in the market
model $\mu\in\mathcal{M}$ \be \boxed{ \label{Xprice} X_\mu(t) =
\mathbb{E}_t^*\left[ X_\mu(T)e^{-\int_t^T
(\alpha^*+\sum_A\alpha^AJ^A_{\mu}) dt' }\right]. \;}\ee
In particular, under the no-arbitrage assumption $\alpha^A = 0$, we recover the classic Martingale pricing theorem:
\be{ X_\mu(t) =
\mathbb{E}_t^*\left[ X_\mu(T)e^{-\int_t^T \alpha^* dt' }\right]\;.}
\ee

}

\bigskip
In section 2.1 we derived a modified Black-Scholes equation for the case of three assets. Now we can use the result of Corollary 3.2 to prove a generalization of such equation. Consider the following vector of assets
\be X = \left[ X_0, X_1,\ldots, X_n, \Phi_1(\vec X, t),\ldots,  \Phi_m(\vec X, t)\right]^\dagger\;.\ee
We will label the components of this vector by $X_\mu$, $\mu = 0,1,\ldots, n+m$.  Moreover, we assume that
$\Phi_i$ are smooth functions of the vector of underlying prices, $\vec X:= [X_1,\ldots, X_n]^\dagger$, and
 $dX_0 = X_0 r dt$  describes a savings account with {\it deterministic} interest rate $r$. The  functions $\Phi_i(\vec X,t)$ describe a set of European-style contingent claims with final payoff $\Phi_i(\vec X(T),T) = f_i(\vec X(T)) $, for some fixed $T\geq t$. Finally, we need to assume that $\alpha^A$ are either deterministic or some function of the underlying prices $\vec X$.
 These assumptions ensure that the expectation values in the RHS of Eq. (\ref{Xprice}) are functions of $\vec X$ and $t$ only, and so our assumption, $\Phi_i = \Phi_i(\vec X,t)$, is self-consistent. Under these assumptions we can prove the following corollary.

 \bigskip

 \noindent {\bf Corollary 3.3 (Modified Black-Scholes Equation)}: {\it Under the assumptions given above, Corollary 3.2 implies that the European-style contingent claims $\Phi_i$,  $i = 1,\ldots, m$, obey the non-linear Black-Scholes equations
\be \boxed{\label{nlBS} {\partial_t \Phi_i +   \sum_{j = 1}^n \left(\alpha^* + \sum_A \alpha^A J^A_j\right) X_j \partial_j \Phi_i - \left(\alpha^* + \sum_A \alpha^A J^A_{n+i} \right)\Phi_i+ \frac{1}{2} \sum_{j,k = 1}^n \Omega_{jk} X_j X_k \partial_j \partial_k \Phi_i = 0\;,} }\\  \ee
with terminal conditions $\Phi_i(\vec X,T) = f_i(\vec X)$.
Moreover,  $J^A = J^A(\vec X,t)$ are a basis for the null space ${\cal N}$  of the $(n+m+1)\times (n+m+1)$ matrix $\Omega$ with components $\Omega_{\mu\nu} = \sum_a \sigma^a_\mu \sigma^a_\nu$, where $\sigma^a_i$, $i = 1,\ldots, n$ are the volatilities of the underlying securities, and we define $\sigma^a_0 := 0$,
\be  \sigma^a_{n+i} := \sum_{j = 1}^n  \sigma^a_j X_j \partial_j \log \Phi_i(\vec X, t)\;,\;\;\; i = 1,\ldots, m\;,\ee
and
\be \label{alphastar} \alpha^* = r  - \sum_A \alpha^A J^A_0  \;.\ee
}

\noindent {\bf Proof:} The Eq. (\ref{nlBS}) of the corollary is a simple application of the Feynman-Kac theorem to Eqs. (\ref{Xprice})  (see \cite{Sh00})). In order to calculate all components of the matrix $\Omega$, we remind the reader that the underlying prices $\vec X$ obey
\be dX_i = X_i \left[ \left(\alpha^* + \sum_A\alpha^AJ^A_i\right)dt + \sum_a \sigma_i^a dW_a^*\right] \;,\;\;\; i = 1,\ldots, n\;.\ee
This implies that the stochastic part of $d\Phi_i$ is given by
\be d\Phi_i(\vec X,t) = \Phi_i(\vec X,t) \sum_{j = 1}^n X_j \partial_j \log \Phi_i(\vec X, t) \sigma^a_j dW_a^* + \ldots \;.\ee
Therefore, the volatilities for the  $X_{n+i} = \Phi_i$ securities are
\be \sigma^a_{n + i} = \sum_{j = 1}^n  \sigma^a_j X_j \partial_j \log \Phi_i(\vec X, t)\;.\ee
Moreover, since $X_0$ is a deterministic process, it follows that $\sigma^a_0 = 0$.
In order to prove Eq. (\ref{alphastar}) of the corollary, we recall that the savings account obey $dX_0 = r X_0 dt$. This implies that $r = \alpha^* + \sum_A \alpha^A J^A_0$. This completes the proof.

\noindent $\square$

The example of section 2.1 is a special case of the corollary 3.3, with $n = m = 1$.   In this case there is only one null direction. We will use the notation $X_1:= X$, $\Phi_1 := V$, $\sigma^1_1 := \sigma_1$, and $\alpha^1:= \tilde \alpha$ in what follows.  A choice for the basis of the null space was given in Eq. (\ref{Jex}), which we repeat here for the convenience of the reader:
\be J =  \frac{1}{\sqrt{2}\sqrt{1 + X\partial X \log V\left( X\partial X \log V - 1\right)} } \begin{pmatrix}
1 - X\partial X \log V \\
X\partial X \log V\\
-1
\end{pmatrix}\;.  \ee
It follows that Eq. (\ref{nlBS}) becomes
\bea 0 &=& \partial_t V +   \left(\alpha^* + \tilde \alpha J_1\right) X \partial_X V- \left(\alpha^* + \tilde \alpha J_2 \right)V +  \frac{1}{2} \sigma_1^2 X^2 \partial_X^2 V \nonumber \\
&=&  \partial_t V +   \left(r + \tilde \alpha (J_1 - J_0) \right) X \partial_X V- \left(r + \tilde \alpha (J_2 - J_0)  \right)V +  \frac{1}{2} \sigma_1^2 X^2 \partial_X^2 V \nonumber \\
&=&   \partial_t V +  r\left( X\partial X V - V\right) +   \frac{1}{2} \sigma_1^2 X^2 \partial_X^2 V \nonumber \\
&&  +  \sqrt{2} \tilde \alpha V  \sqrt{1 + X\partial X \log V\left( X\partial X \log V - 1\right)}\;.\eea
This is exactly what we obtained in section 2.1 (c.f. Eq. (\ref{BS})).

\subsection{Relation to Farinelli Connection}

Before concluding this section, we would like to relate our
connection $\Gamma$ to another arbitrage connection proposed
recently in \cite{Far08}, \cite{Far09a} and \cite{Far09b}. We will
show that the connection presented in \cite{Far08}, \cite{Far09a}
and \cite{Far09b} is equivalent to Eq. (\ref{gamma}). Consider an
economy in the time interval $t \in [0,T]$, composed of some general
``base assets" with prices $X_i$, $i=1,\ldots,N$, and synthetic
zero-bonds on these assets. The price of a zero-bond (in the same
units as $X_i$) can be defined as \be Z_i(t,T) := B_i(t,T) X_i(t)\;,
\;\;\; i = 1,\ldots, N\;, \ee where $B_i(T,T) = 1$ and $B(t,T)>0$,
if $T\ge t$. In other words, $Z_i$ pays one unit of asset $X_i$ at
maturity. Moreover, the dynamics of $B_i(t,T)$ can be traced back to
a $T$-independent stochastic, which we call the {\it spot rate}
$r_i$. Explicitly, \be\label{BF} r_i(t):=\lim_{t\rightarrow
t^-}-\frac{\partial}{\partial T}\log B_i(t,T)\text{  or equivalently
} B_i(t,T) := \mathbb{E}_t\left[e^{-\int_t^T r_i(s)ds}\right]\;.\ee
The following gauge connection was proposed for this
economy\footnote{In \cite{Far08}, \cite{Far09a} and \cite{Far09b},
Stratonovich calculus was used instead of It\^o \cite{VKr81}. Here
the usual differentiation rules apply.  }
\bea \label{F} F &=& \frac{\sum_i X_i \left( d\phi_i - r_i \phi_i  dt\right)}{\sum_j \phi_j X_j} \nonumber \\
&=& d\log\left(\sum_i  \phi_i X_i\right) - \frac{\sum_i  \phi_i X_i \left(d \log X_i  + r_i dt\right)}{\sum_j \phi_j X_j}\;.\eea
Our goal is to relate $F$ to our connection $\Gamma$, Eq. (\ref{gamma}), restricted only to the base assets.

In order to do this, we define a larger price vector with components
$X_\mu$, $\mu = 1,\ldots, 2N$: \be X(t) = \left( X_1(t),\ldots,
X_N(t), Z_1(t,T),\ldots , Z_N(t,T)\right)^{\dagger}\;.\ee We now do
the usual decomposition of $dX_\mu$ as in Eqs. (\ref{Ito}) and
(\ref{alphamu}). Using the pricing formula, Eq. (\ref{Xprice}), we find that
\bea B_i(t,T)  &=& \mathbb{E}_t^*\left[ e^{- \int_t^T (\alpha^* + \sum_A \alpha^A J^A_{i+N}) dt' + \int_t^T d\log X_i}\right] \nonumber \\
&=& \mathbb{E}_t\left[ e^{- \int_t^T \left[ (\alpha^* + \sum_A \alpha^A J^A_{i+N} + \sum_a(\beta^a)^2/2)dt' +\sum_a \beta^a dW_a -  d\log X_i\right]  } \right]\nonumber \\
&=& \mathbb{E}_t\left[ e^{-\int_t^T r_i(s) ds} \right]\;,\eea where
we have extracted the spot rate as \be \label{rit}  r_i = \alpha^* +
\sum_A \alpha^A J^A_{i+N} +\frac{1}{2} \sum_a(\beta^a)^2 +\sum_a
\beta^a \frac{dW_a}{dt} - \frac{d}{dt}\log X_i\;.\ee  Moreover,
since $\lim_{T \rightarrow t^+} Z_i(t,T) = X_i(t)$,  and by assumption $r_i$ is independent of $T$, it must be that
\be J^A_{i+N} = J^A_{i}\;.\ee Therefore, we
can finally identify the spot rate as: \be \label{ri} r_i = \alpha^* +
\sum_A \alpha^A J^A_{i} +\frac{1}{2} \sum_a(\beta^a)^2 +\sum_a
\beta^a \frac{dW_a}{dt} - \frac{d}{dt}\log X_i\;.\ee Inserting Eq.
(\ref{ri}) in (\ref{F}) we obtain \be F =  - \frac{\sum_{i, A}
\alpha^A J^A_i \phi_i X_i }{\sum_j \phi_j X_j} dt + d\Lambda\;, \ee
where \be d\Lambda = -\left( \alpha^* + \frac{1}{2} \sum_a (\beta^a)^2
\right)dt - \sum_a \beta^a dW_a + d\log\left(\sum_i \phi_i
X_i\right) \;.\ee Therefore, we conclude that the connection $F$ is
equivalent to $\Gamma$ up to a sign and a gauge transformation.

\section{Measuring Arbitrage Curvature}
In this section we explain how to estimate the arbitrage parameters
$\alpha^A$ using financial data. Given the discussion in the
previous section, measuring these parameters is equivalent to
measuring the ``curvature"  of the market. Needless to say, one
can do this for a subset of all instruments only, and there are many
technical difficulties which we discuss below.

Even though $\alpha^A$ is a gauge invariant,  it is still defined up
to a rotation in the null space\footnote{For notational simplicity,
we will still use ${\cal N}$ for the null space of the particular
market sub-sector under study. However, it is important to keep in
mind that this is not the null space of the full market. }  ${\cal
N}$. Therefore, the basic idea is to measure the rotational and
gauge invariant quantity \be\label{alphasquare} \sum_{\mu, A}
\frac{\alpha^A J^A_\mu}{X_\mu} \frac{dX_\mu}{dt} = \sum_A
(\alpha^A)^2   \geq 0\;,\ee where $\alpha^A$ in the left hand side
of this equation is expressed as \be \alpha^A = \sum_\mu
\frac{J^A_\mu}{X_\mu} \frac{dX_\mu}{dt}\;.\ee The vectors $J^A$ must
be calculated using an estimate for the quadratic variation
$\Omega_{\mu \nu} := d\bra \log X_\mu, \log X_\nu\ket/dt$, and the
results of Proposition 2.2. We introduce the notation
$\mathcal{A}^2:=\sum_A(\alpha^A)^2$ for the measurement of
arbitrage curvature. A positive detection of ${\cal A}^2$ can be translated into a self-financing arbitrage portfolio strategy using the result of the following proposition.

\bigskip
\noindent {\bf Proposition 4.1 (Arbitrage Strategy)}\label{Prop41}:
{\it Let the asset corresponding to $\mu=0$ be the num\'{e}raire ($X_0 :=1$). If
the market model satisfy the positive curvature assumption
\begin{equation}
\mathcal{A}^2>0,
\end{equation}
then the portfolio allocation
\begin{equation}
\begin{split}
\phi_{0}(t)&:=\sum_{i=1}^N\int_0^t\phi_i(s)dX_i(s)+\sum_A J^A_{0}(t)\alpha^A(t)\;, \\
\phi_{i}(t)&:=\sum_A \frac{J^A_{i}(t)\alpha^A(t)}{X_{i}(t)}\quad(i=1,\dots,N)\;,
\end{split}
\end{equation}
is a selfinancing arbitrage strategy delivering wealth
\begin{equation}
V(t)=\int_0^t\mathcal{A}^2(s)ds.
\end{equation}
}

\noindent {\bf Proof:} First, we check that the strategy is
self-financing, that is
\begin{equation}
\sum_{\mu=0}^N\left( d\phi_{\mu}X_{\mu}+d\bra \phi_{\mu}, X_{\mu}\ket \right) =0 \;,
\end{equation}
where $d$ denotes the It\^{o} differential (see \cite{LaLa06},
Chapter 4.1.2). This is proved by the following computation:
\bea
\sum_{\mu=0}^N \left( d\phi_{\mu}X_{\mu}+d \bra \phi_{\mu}, X_{\mu}\ket\right)  &=& d\phi_{0}X_{0}+d\bra \phi_{0},X_{0} \ket + \sum_{i=1}^N \left( d\phi_{i}X_i +d\bra\phi_{i},X_{i}\ket \right)\nonumber \\
&=& \sum_A \sum_{i=1}^N \alpha^A J^A_i{dX_i}/{X_i}  +  \sum_A d\left( \alpha^A J^A_0 \right) \nonumber \\
&& + \sum_{i = 1}^N X_i d\left(  \sum_A  \alpha^A{J^A_{i}}/{X_{i}}\right)  + \sum_{i  =1}^N  d \bra  \sum_A  \alpha^A{J^A_{i}}/{X_{i}} , X_i \ket \nonumber \\
&=& \sum_A  d\left( \alpha^A  \sum_{\mu = 0}^N J^A_\mu\right)\nonumber \\
&=& 0\;.\eea
Since the self-financing condition is fulfilled, the portfolio value
can be computed as

\bea
V(t)&=&\sum_{\mu=0}^N\phi_{\mu}(t)X_{\mu}(t) =\int_0^t\phi_{i}(s)dX_{i}(s) + \sum_A \alpha^A\underbrace{\sum_{\mu=0}^NJ^A_{\mu}}_{=0} \nonumber\\
&=&\int_0^t\sum_A \alpha^A(s)\sum_{i=1}^N\frac{J^A_{i}(s)}{X_{i}(s)}dX_i(s)= \int_0^t\sum_A \alpha^A(s)\sum_{\mu=0}^N\frac{J^A_{\mu}(s)}{X_{\mu}(s)}dX_\mu(s)\nonumber \\
&=& \int_0^t\sum_A\alpha^A(s)^2ds =\int_0^t\mathcal{A}^2(s)ds.
\eea
Since the arbitrage curvature are positive, we see that $V(0)=0$ and
$V(t)>0$ for all times $t\in[0,T]$. The proof is completed.

 \noindent {$\square$}

Of course, there is no continuous time trading in the markets, and
we can only do measurements in discrete time. Moreover, our estimate
of $\Omega$ will always include errors. This means that we will
always have a noise term in the right hand side of Eq.
(\ref{alphasquare}). The goal of this section is to explain the
basic steps used to measure arbitrage curvature, and understand the
major sources of error in such measurements. A key aspect of our
algorithm is that we test directly for the gauge invariance of the
arbitrage signal. This allow us to check the robustness of our
estimators. We find that the gauge invariance of the arbitrage
signal, as predicted by the stochastic models, is indeed obeyed
with good accuracy in the real market.

\subsection{Basic Algorithm}
In what follows, we will use a hat in any variable which is an
estimate of some parameter, e.g. $\hat \Omega$ is an estimate for
$\Omega$. The first problem we face is to find an estimate for the
quadratic variation $\Omega$ and to determine the null space ${\cal
N}$ defined in section 2 (if non-trivial). This is a familiar
problem in volatility modeling.  Since we will never observe
$\Omega$ directly, it is expected that our estimate will not have
any exact zero-mode, but only eigenvectors with small eigenvalues.
In fact, a priori, we do not know if the space ${\cal N}$ is
non-trivial. We can only guess its dimension.

Let $\hat \Omega$ be any estimate for $\Omega$. Then, following
Proposition 2.2, we construct the matrix \be \label{Ghat} \hat G =
\hat \Omega  - \frac{1}{N} \left( U \hat \Omega + \hat \Omega
U\right) + \frac{1}{N^2} \tr(U \hat \Omega) U\;,\ee where $N$ is the
number of rows (or columns) of $\Omega$ and $U$ is the matrix of all
ones: $U_{\mu\nu} = 1\;, \; \forall \mu,\nu$. We can then use
standard algorithms to compute the eigenspace of $\hat G$. This will
yield orthonormal eigenvectors \be \hat G \hat J^A = \lambda^A \hat
J^A\;,\;\;\; (\hat J^A)^{\dagger} \hat J^B = \delta^{A B}\;,\;\;\;
A,B = 0,1,\ldots, N-1\;,\ee where $\lambda^A \geq 0$ since $\hat G$
is positive semidefinite. As a matter of fact, since $\Omega$ and
$U$ commute, the have a common basis of eigenvectors and a short
computation proves that $G$ has always (at least) one zero
eigenvalue and the biggest $N-1$ eigenvalues equal those of
$\Omega$, which are not negative  (c.f. Proposition \ref{Prop22}.2).
In practice, there will only be one exact zero eigenvector: $\hat
J^0 \propto (1,1,\ldots,1)^{\dagger}$. Summarizing, the eigenvalues
of $G$ ordered in increasing order of magnitude are, \be 0 =
\lambda^0 \leq \lambda^1 \leq \lambda^2\leq \ldots \leq
\lambda^{N-1}\;.\ee It is easy to show (c.f. Proposition 2.2) that
\be \sum_\mu \hat J^A_\mu = 0\;,\;\;\; \text{for}\;\; A=1,2,\ldots,
N-1\;.\ee Our estimate for the basis of ${\cal N}$ will be to chose
the first $k$ eigenvectors with the smallest eigenvalues: $\hat
J^A$, $A = 1,\ldots, k < N-1$. In doing this, we are assuming that
$\text{dim}({\cal N}) = k$.

Once, we have calculated  $\hat J^A$, we can compute our estimate of
$\alpha^A$  in discrete time: \be\label{alphahat}
\boxed{\hat{\alpha}^A(t + \delta t) =\sum_\mu \frac{\hat
J^A_\mu(t)}{\delta t X_\mu(t)} \left[X_\mu(t+\delta t) -
X_\mu(t)\right]\;.}\ee Note that $\hat J^A(t)$ is constructed with
information up to time $t$ only. This estimate is consistent with
the non-anticipating nature of It\^o integrals. The time step
$\delta t$ is, of course, arbitrary.  Our estimate for
$\mathcal{A}^2$ now becomes: \be \label{alphahatsquare}
\boxed{\hat{\mathcal{A}^2}(t + \delta t) = \sum_{A = 1}^k \left[\hat
\alpha^A(t)\right]^2 + \sum_{A = 1}^k \hat \alpha^A(t)\left[ \hat
\alpha^A(t+\delta t) - \hat \alpha^A(t)\right]\;.}\ee In the limit
of short time scales, and if there is non-trivial arbitrage, we
expect that this estimator converges to the true signal \be
\label{limit}  \hat{\mathcal{A}^2}(t + \delta t) = \mathcal{A}^2(t)
+ \sum_A \alpha^A d\alpha^A = \mathcal{A}^2(t) + {\cal O}(\delta
t)\quad(\delta t \rightarrow 0).\ee The convergence in Eq.
(\ref{limit}) is only valid if in the limit $\delta t \rightarrow 0$
we have \be \mathbb{E}_t\left[ \hat \alpha^A(t+\delta t)\right]-
\hat \alpha^A(t) = {\cal O}(\delta t)\;, \;\;\; \text{Cov}_t\left[
\hat \alpha^A(t+\delta t),  \hat \alpha^B(t + \delta t)\right] ={\cal O}( \delta
t)\;.\ee Therefore, we expect that, if there is non-trivial
arbitrage in the market, the estimator (\ref{alphahatsquare}) will
give us a positive signal on average. Since the time scale is
arbitrary, it is convenient to set $\delta t = 1$ henceforth.

There are several candidates for an estimator for $\Omega$. The
``right" choice of $\hat{\Omega}$ should reflect our believes about
the true dynamics of the asset values. Here we will simply take the
empirical estimator for covariance of the time series of log returns
for a window of length $L$. More precisely, our data consist of a
number of time series for the prices $X_\mu$, $\mu = 0, \ldots, N-1$
in certain units, say USD\footnote{We also include the USD itself as
an asset in which we have $X_0 = 1$.}. Our estimator reads \bea
\label{Cov} \hat \Omega_{\mu\nu} (t)
&=& \frac{1}{L} \sum_{i = 0}^{L-1} \log \left[ \frac{X_\mu(t-i)}{X_\mu(t-i-1)}\right]\log \left[ \frac{X_\nu(t-i)}{X_\nu(t-i-1)}\right] \nonumber \\
&& - \frac{1}{L^2} \sum_{i,j = 0}^{L-1} \log \left[
\frac{X_\mu(t-i)}{X_\mu(t-i-1)}\right]\log \left[
\frac{X_\nu(t-j)}{X_\nu(t-j-1)}\right]\;.  \eea For more
sophisticated estimators see \cite{HaHaPi08, Zh06}. We are now in position
to summarize the most basic algorithm to detect arbitrage.

\noindent
{\bf Algorithm:}
{\it
\begin{enumerate}
\item Starting with the time series for $X_\mu$, $\mu = 0,\ldots, N-1$ in an interval $[t, t-L]$, we estimate $\hat \Omega_{\mu\nu}(t)$ using Eq. (\ref{Cov}).
\item We then calculate the $\hat G$ matrix using Eq. (\ref{Ghat}), and its orthonormal eigenspace. The eigenvectors will be labeled as $\hat J^A$, $A = 0,1,\ldots, N-1$, in order of increasing eigenvalues: $0 = \lambda^0 \leq \lambda^1\leq \ldots \leq \lambda^{N-1}$. Moreover, $\hat J^0 \propto (1,1,\ldots,1)^{\dagger}$.
\item Given a guess for the dimension of the null space $k  = \text{dim}({\cal N})$, we take as its basis the following eigenvectors of $\hat G$:  $\hat J^A$, $A = 1,\ldots, k$.
\item We then calculate $\hat \alpha^A(t+1)$ from Eq. (\ref{alphahat}), which uses information up to time $t+1$.
\item Roll the time window by one step, and repeat steps 1-5. Once we have more than one estimate for $\hat \alpha^A$ , we can calculate our final arbitrage estimator $\hat{\mathcal{A}}^2$ from Eq. (\ref{alphahatsquare}).
\item In order to explicitly check for gauge invariance, we repeat steps 1-5, using each asset $X_\mu$ as numeraire. For example, if we want to use $X_1$ as numeraire, we divide all elements of the time series by the corresponding element of $X_1$, e.g.  $X_\mu(s) \rightarrow X_\mu(s)/X_1(s)$  $\forall \mu$ and $s \in [t,t-L]$. Then we repeat steps 1-5 with the new time series. Note that this is a non-trivial transformation in the data and, in practice, we will get different estimates for $\hat\alpha^A$.
\end{enumerate}
}

Before discussing the results of the algorithm, we need to understand what are the main sources of error in our signal. This is done in the next subsection.

\subsection{Sources of Error}
The sources of error in our measurement of ${\cal A}^2$ can be divided in three groups. First, there
is gauge dependence. Second, there is a gauge invariant noise, which
we will discuss below. Finally, when using high-frequency financial data, one is faced with the so-called market microstructure noise which is partly due to the bid/ask bounce effect \cite{HaHaPi08}.

We begin with looking at sources of gauge
dependence. Note that our construction of the estimators {\it
assumes} that, under a gauge transformation,  $\hat \Omega$
transforms like $\Omega$, c.f. Eq. (\ref{Omegaxform}). However, the
gauge transformation rule in the real world can be quite different,
because the unknown effective dynamics could lead to gauge
dependences. We do not have an {\it a priori} test for this source
of error. The only way to test for it is to make our calculations in
different gauges and see how different the answers are. We will show
examples of this in the next sections.

The second  source of error in our
signals come from a gauge invariant noise term. In fact, we will see that this is the dominant noise contribution. In order to
understand this noise, it is convenient to discretize the It\^o
integral and write our estimate for $\alpha^A$ as
 \bea \hat\alpha^A(t+1) &=& \sum_\mu\frac{ \hat J^A_\mu(t)}{X_\mu(t)} \left[ X_\mu(t+1) - X_\mu(t)\right] \nonumber \\
 &=& \sum_{\mu, B} \hat J^A_\mu(t) J^B_\mu(t) \alpha^B(t) + \sum_{\mu,a} \hat J^A_\mu(t) \sigma^a_\mu \beta^a(t)  \nonumber \\
 && + \sum_{\mu,a} \hat J^A_\mu(t) \sigma^a_\mu \left[W_a(t+1) - W_a(t)\right] \nonumber \\
 &:=& \alpha^A_\text{trend}(t) + \epsilon^A(t+1)\;.\eea
 Here we have decomposed the signal in a trend
 \be \alpha^A_\text{trend}(t)  := \sum_{\mu, B} \hat J^A_\mu(t) J^B_\mu(t) \alpha^B(t) + \sum_{\mu,a} \hat J^A_\mu(t) \sigma^a_\mu \beta^a(t)\;,\ee
 and a stochastic noise term
 \be \epsilon^A(t+1):=  \sum_{\mu,a} \hat J^A_\mu(t) \sigma^a_\mu \left[W_a(t+1) - W_a(t)\right]\;,\ee
with $\mathbb{E}_t[\epsilon^A(t+1)] = 0$. Since $\hat J^A$ is only
an estimate for the real $J^A$, we have that $\sum_\mu \hat J^A
\sigma^a_\mu \neq 0$ in general. Therefore, our error in the
estimate of $J^A$ will induce an extra noise term in the signal.
Moreover, it will also induce some gauge dependency. To see this,
note that under a change of num\'{e}raire, we have $\sigma^a_\mu
\rightarrow \sigma^a_\mu + \delta \sigma^a$ and $\beta^a \rightarrow
\beta^a + \delta\sigma^a$. It is then easy to check that the trend
will transform as $\alpha^A_\text{trend} \rightarrow
\alpha^A_\text{trend} + \sum_{\mu,a} \hat J^A_\mu \sigma^a_\mu
\delta\sigma^a$.  However, note that the noise term is gauge
invariant.  In fact, one expects the term $\sum_\mu \hat J^A_\mu
\sigma^a_\mu$ to be quite small. Moreover, since in our algorithm
the gauge transformation is of the order
$\delta\sigma^a={\cal O}(\sigma^a_\mu)$, we expect the gauge dependence
coming from the trend to be negligible.  We will see that, in real
financial data, most of the signal can be accounted by the gauge
invariant noise term.\par We are interested in estimating the size
of the noise contribution. For that, we compute the variance of the
noise using information up to time $t$:
 \bea \text{Var}_t\left[ \sum_A \hat \alpha^A(t) \hat \alpha^A(t+1)\right]  &=& \mathbb{E}_t\left\{ \left[ \sum_A \hat \alpha^A(t) \left(\hat \alpha^A(t+1) - \mathbb{E}_t[\hat \alpha^A(t+1)]\right)\right]^2\right\} \nonumber \\
 &=& \sum_{A,B} \hat  \alpha^A(t) \hat \alpha^B(t)\left( [\hat J^A(t)]^{\dagger}  G(t)   \hat J^B(t)\right)\nonumber \\
 &=& \sum_A [\hat \alpha^A(t)]^2 \lambda^A(t) \nonumber \\
 &&+ \sum_{A,B} \hat  \alpha^A(t) \hat \alpha^B(t)\left( [\hat J^A(t)]^{\dagger}  \delta G(t)   \hat J^B(t)\right)\;, \eea
 where $\delta G = G- \hat G$. If we think that our estimate of $G$ is good, we can neglect the $\mathcal{O}(\delta G)$ term and approximate
 \be  \label{error} \boxed{\text{Var}_t\left[ \hat{\mathcal{A}}^2(t+1)\right] \approx \sum_A [\hat \alpha^A(t)]^2 \lambda^A(t) \leq \sum_A [\hat \alpha^A(t)]^2 \lambda_k(t)\;,}\ee
where we remind the reader that $k$ is our estimate for the
dimension of ${\cal N}$, and the eigenvalues of $\hat G$ have been
ordered so that $\lambda^1 \leq \lambda^2\leq \ldots \leq
\lambda^k$. \par An interesting consequence of Eq. (\ref{error}) is
that one can put a fundamental bound for the size of the arbitrage
curvature, in order to be detectable. We have, \be \mathcal{A}^2
\geq \sqrt{\text{Var}_t\left[ \hat{\mathcal{A}}^2\right]} \implies
\mathcal{A}^2 \geq \lambda_k\;.\ee This means that, in order to have
a chance to detect arbitrage, one needs to find financial products
whose time series are as correlated as possible, which implies a
very small value of $\lambda_k$.

The third source of error is market microstructure noise. This effect is relevant in high frequency data, when the size of the price movements is comparable with the bid/ask spread. In order to model this noise, it is convenient to set $X_0: =1$ as our num\'eraire. The standard way of simulating this noise is to introduce an additional jump term $\eta_i(t)$ to the log prices $X_i$, $i = 1,\ldots, N$. More precisely, the observed price is $\tilde X_i$ and it is given by,
\be \label{Logcont} \log \tilde X_i(t) = \log X_i(t) + \eta_i(t)\;,\ee
where $X_i$ is the ``true" It\^o process, and for simplicity we assume
\be \mathbb{E}[X_i \eta_j] = 0\;,\;\;\;  \mathbb{E}[\eta_i] = 0\;,\;\;\; \mathbb{E}[\eta_i \eta_j] := {\eta^2} \delta_{i j} \;.\ee
Moreover, the noise terms are uncorrelated between different times.
One can then show that our estimator will be contaminated by an amount
\be \label{Acont} \lim_{\delta t \rightarrow 0} \mathbb{E}[\hat {\cal A}^2] = \mathbb{E}[{\cal A}^2] - \gamma\frac{\eta^2}{\delta t^2} + {\cal O}\left(\frac{\eta^4}{\delta t^2}\right)\;,\ee
where $\gamma := \text{dim}({\cal N}) - \mathbb{E}[\left( \sum_{A} J^A_0\right)^2] $.  It  can be shown that $\gamma \geq 0$.  Therefore, we see that the microstructure noise leads a negative contribution to our estimation of ${\cal A}^2$.  The absolute value of such contribution diverges as we move towards higher frequencies ($\delta t \rightarrow 0$).

One way of detecting the presence of microstructure noise is to note that
\be \label{Logcorr} \lim_{\delta t \rightarrow 0}  \mathbb{E}\left[\log\left(\tilde X_i(t+\delta t)/\tilde X_i(t)\right)\log\left(\tilde X_i(t)/\tilde  X_i(t-\delta t)\right)\right]  = - \eta^2 < 0\;.\ee
In other words, the microstructure noise induces a negative correlation between subsequent log returns. We find that this effect is quite pronounced for equity and futures data. However, for stock indices, the effect seems to be negligible. This is mainly due to the fact that the microstructure noise ``averages out" between all the stocks in the index.

There is an extra source of error which is intrinsic to the
algorithm, but only if we use a rolling window in our estimation of
$\hat J^A$. For example, suppose that we estimate $\hat J^A(t)$ and
then roll the window and estimate $\hat J^A(t+1)$. Even if the
matrices $\hat G(t)$ and $\hat G(t+1)$ are near, $\hat J^A(t)$ and
$\hat J^A(t+1)$ can differ by a large orthogonal transform. It can
be just a sign flip f.i., since the eigenvalue equations are
invariant under $\hat J^A \rightarrow - \hat J^A$. However, suppose
two eigenvalues are near to each other, i.e. $\lambda_1 \approx
\lambda_2$. Then, any linear combination of $\hat J^1$ and $\hat
J^2$ is also approximatively an eigenvector of $\hat G$.  In
physics, this is known as the problem of {\it degenerate
perturbation theory} (see f.i. \cite{Sa94}). More generally, we have
that \be \lim_{|| \hat G(t) - \hat G(t+1)|| \rightarrow 0} \hat
J^A(t+1)= \sum_{B}  C^{AB} \hat J^B(t)\;,\ee where $C$ is an orthogonal matrix, i.e.
\be C^{\dagger} C = \mathbf{1}\;.\ee The problem can be solved if we
can determine $C$. If so,  we can construct the ``correct"
eigenvectors $\tilde J^A(t+1):=  \sum_B (C^\dagger)^{AB} \hat J^B(t+1) $ so that $\lim_{|| \delta
\hat G||\rightarrow 0 } \tilde J^A(t+1) = \hat J^A(t)$.  An approximate
solution for $C$ is presented in appendix A. This is implemented in
our numerical routines.

\subsection{Simulated Data}
In this section we apply our algorithm to  simulated financial data.
For this, we study the simple log-normal random walk model with
constant coefficients (c.f. Eq. (\ref{Ito})). The solution to the
stochastic differential equation (\ref{Ito}) is \be \label{Itosim}
X_\mu(t) = X_\mu(0)\exp \left[\left( \alpha_\mu - \frac{1}{2}
\sum_{a = 1}^d(\sigma^a_\mu)^2\right) t + \sum_{a = 1}^d
\sigma^a_\mu B_a(t)\right]\;,\ee
  where $B(t):=[B_1(t),\dots,B_d(t)]^{\dagger}$ is standard multivariate Brownian motion with
  \be \text{E}[B_a(t)] = 0\;,\;\;\;  \text{Cov}[B_a(t),B_b(t)] =  t \delta_{ab}\;,\ee
for all $a,b=1,\dots,d$. As usual, we decompose the trends as
  \be\alpha_\mu = \alpha + \sum_{a=1}^d \beta^a \hat\sigma_\mu^a  + \sum_{A \in \cal N}  \alpha^A J^A_\mu \;,\;\;\;  \hat \sigma^a_\mu =  \sigma^a_\mu - \frac{1}{N} \sum_{\nu=0}^{N-1} \sigma^a_\nu\;.\ee
We begin with an example with $N = 21$ assets and $d = 18$ Brownian
motions, which implies $k = \text{dim}({\cal N}) = 2$.  We take as a
first asset a bank account with zero interest rate, and make it to our num\'{e}raire.
This means that we choose, \be X_0 :\equiv 1\;,\;\;\;\sigma^a_0 =
0\;,\;\;\; \alpha_0 = 0\;,\ee which implies \be \alpha = \frac{1}{N}
\sum_{i = 1}^{N-1} \sum_{a = 1}^d \beta^a\sigma_i^a - \sum_{i =
1}^{N-1} \sum_{A =1}^2 \alpha^A J^A_i\;.\ee In figure 1 we show a
particular simulation of the log prices, where we take
$\beta^a,\sigma^a_i,\alpha^A$ from uniform random distributions in
the intervals, $\beta^a \in [-10^{-4},10^{-4}]$, $\sigma^a_i \in
[-10^{-3},10^{-3}]$ and $\alpha^A \in [-10^{-4},10^{-4}]$. The
simulation was generated using Mathematica. The arbitrage detection
algorithm was implemented in C++. Each price was taken at a time
separation  of $\Delta t = 1 \text{ (arbitrary time unit)}$. In this
particular case we calculate $\hat \Omega$ using the first 100
prices of the time series. In other words, we do not use a moving
window. The results with the moving window are very similar.
\myfig{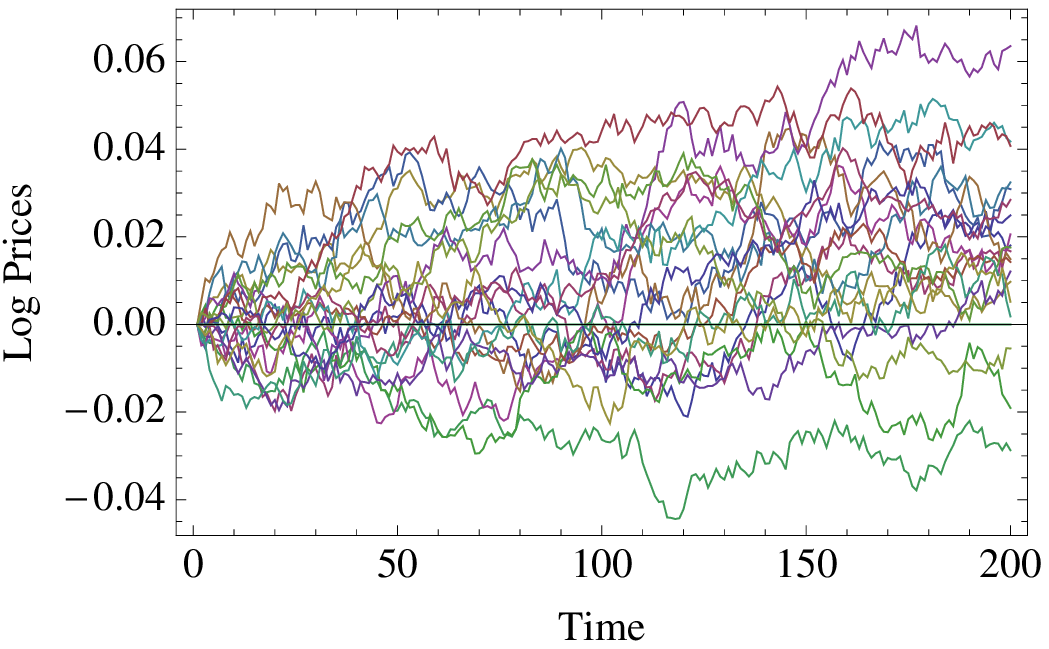}{12}{\small \it Simulation with 20
log-normal random  walks. }

Now suppose we assume (correctly) that we have $k = \text{dim}({\cal
N}) = 2$. We then run the algorithm and find the signal shown in
figure 2. The solid horizontal line at $\mathcal{A}^2 \approx
10^{-8}$ is the correct value of $\mathcal{A}^2$. Therefore, we see
that we get an accurate estimate for the arbitrage curvature. Note
that, as we discussed in the previous section, in our algorithm we
compute $\hat{\mathcal{A}}^2$ using each of the different assets as
num\'{e}raire. We include error bars showing the range of values
obtained  using the different gauges. The results in this simulated
sample are gauge invariant to such high accuracy that the error bars
cannot be appreciated.\par In the previous section we discussed how
the main source of error in our detection technique can be related
to the biggest eigenvalue $\lambda_k$ of the set
$\{\lambda_1,\ldots,\lambda_k\}$. This led to a gauge invariant
noise term.  In this simulated sample data, we find that $\lambda_k
\approx 10^{-21}$, and so using Eq. (\ref{error}) we find,
$\sqrt{\text{Var}[\hat{\mathcal{A}}^2] }\approx \sqrt{10^{-8}\times
10^{-21}}\approx 10^{-15}$. Therefore, this noise term is very small in
this case. The fluctuations seen in figure 2 are an artifact of this
particular model. To understand them, we can expand Eq.
(\ref{Itosim}) as
  \be  \frac{ X_\mu(t + 1) - X_\mu(t)}{  X_\mu(t)} = \alpha_\mu  +\sum_a \sigma^a_\mu B_a(1)+ \epsilon_\mu +\ldots\;,  \ee
where \be \epsilon_\mu  = \frac{1}{2} \left( -\Omega_{\mu\mu} +
\sum_{a,b} \sigma^a_\mu \sigma^b_\mu B_a(1) B_b(1) \right)\;.\ee It
is easy to show that $\epsilon_\mu$ is gauge invariant and \be
\mathbb{E}[\epsilon_\mu] = 0\;,\;\;\; \mathbb{E}[\epsilon_\mu
\epsilon_\nu] = \frac{1}{2} (\Omega_{\mu\nu})^2\;.\ee This extra
noise term, $\epsilon_\mu$, is the reason for the gauge invariant
fluctuations in figure 2. The noise term vanishes, if we integrate
$dX_\mu$ using an infinite partition of the time interval, as it is
assumed in It\^o integrals. Of course, this is never possible in
practice. Nevertheless, we see that in this example, the extra noise
is very small compared to the arbitrage parameter $\mathcal{A}^2$.
In fact, we expect this noise to be very small in general since it
is of order $\text{Var}[\epsilon_\mu]=O((\sigma^a_\mu)^4)$.\par It
is interesting to see what happens if we assume the wrong number of
zero modes. For example, in figure 3 we show what happens if we take
$k=1$. We see that we get a gauge dependent signal. Finally, in
figure 4 we show what happens if we assume $k = 3$. In this case,
the biggest eigenvalue is $\lambda_k \approx 10^{-8}$. As the figure
shows, most of the fluctuations are coming from the gauge invariant
noise described in the previous section. To see this we have plotted
the expected  noise according to Eq. (\ref{error}): \be
\label{noiseest} \text{noise}_\pm(t+1) = \left(\mathcal{A}^2 \pm
\sqrt{\text{Var}_t[\hat{\mathcal{A}}^2(t+1)]}\right) = \left(
\mathcal{A}^2 \pm \sqrt{\sum_{A=1}^k (\hat\alpha^A(t))^2
\lambda^A(t)} \right)\;,\ee where $\mathcal{A}^2 \approx 10^{-8}$ is
the true value of the arbitrage (which is also the mean of the
signal). We see that this noise accounts for most of the
fluctuations and it makes the true arbitrage signal almost undetectable. The main point we would like to make here is that the correct value
of $k$ can be estimated from the quality of the signal.

\myfig{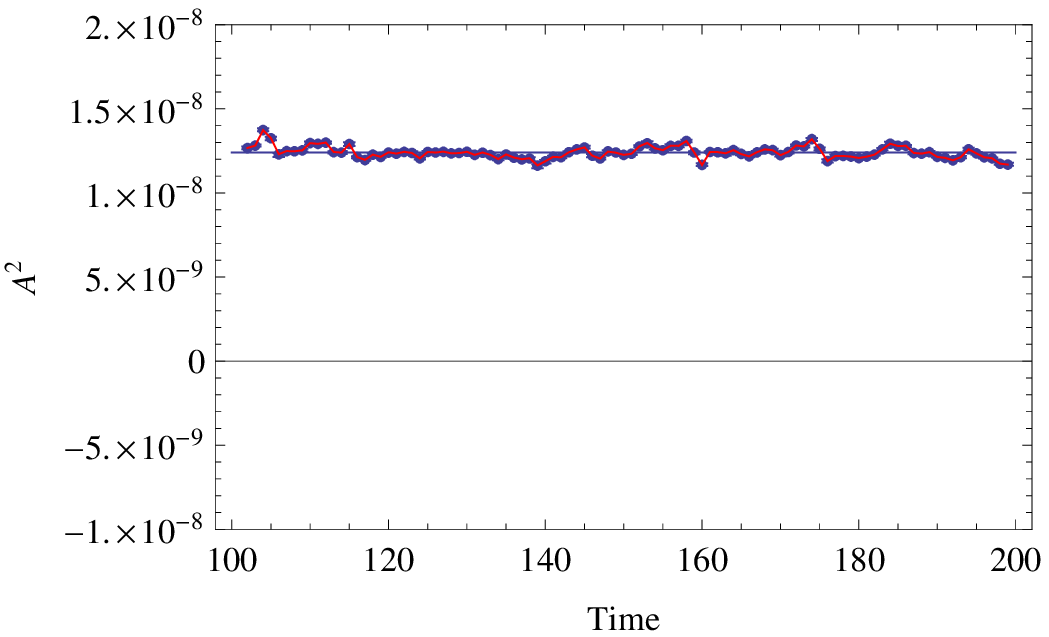}{12}{\small \it Result for the arbitrage
detection algorithm applied to the simulated data in figure 1.  Here
we assume (correctly) $k = 2$ null directions. The solid horizontal
line at ${\cal A}^2 \sim 10^{-8}$ is the correct value of ${\cal A}^2$.
The red solid line is the USD value of the signal.  }
\myfig{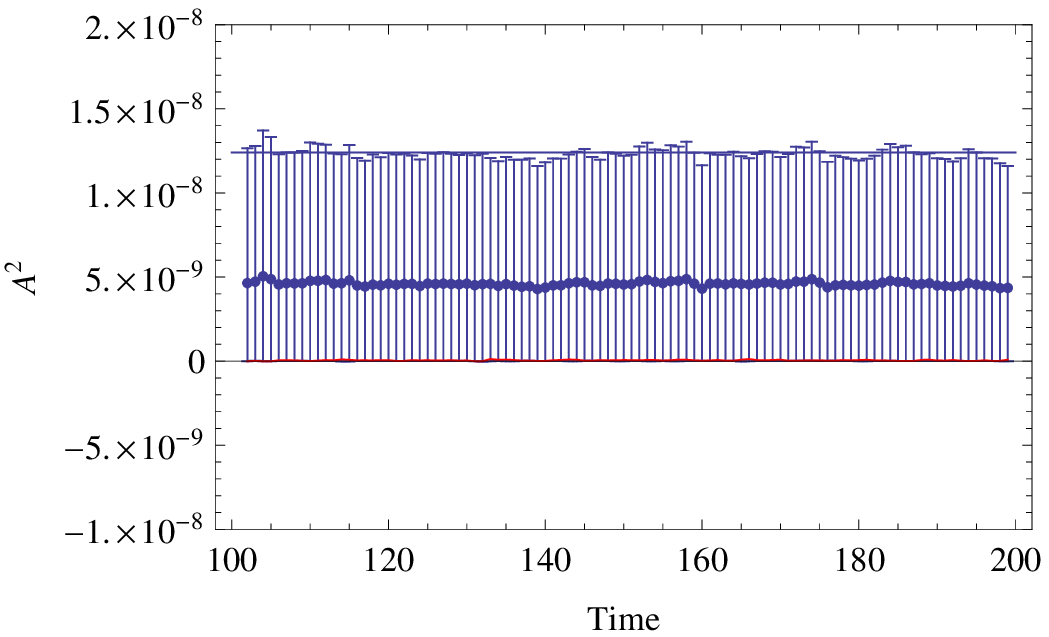}{14}{\small \it Result for the arbitrage
detection algorithm applied to the simulated data in figure 1. Assuming $k = 1$ null directions (incorrectly), and using
a fixed window of 100 time steps. The error bars give the range of
values obtained using the different gauges. The solid dots are the
mean of all results. The red line is the USD signal.  }
\myfig{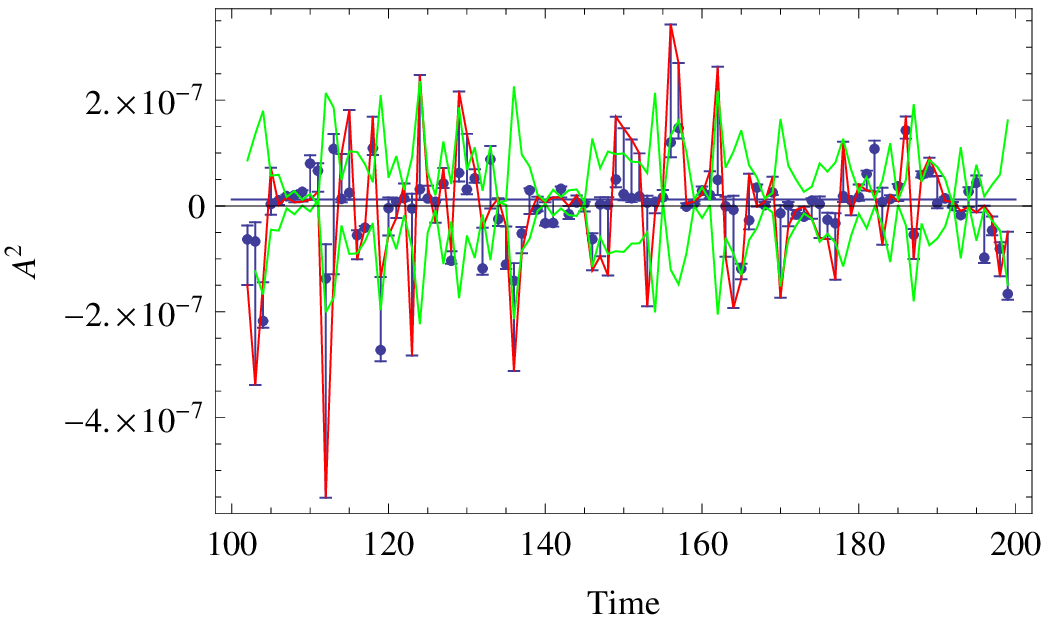}{14}{\small \it Result for the arbitrage
detection algorithm applied to the simulated data in figure 1. Assuming $k = 3$ null directions (incorrectly), and using
a fixed window of 100 time steps. The green lines are the noise
terms $\text{noise}_\pm$ estimated according to Eq.
(\ref{noiseest}). The error bars give the range of values obtained
using the different gauges. The solid dots are the mean of all
results. The red line is the USD signal. }

We can now investigate the effect of the market microstructure noise discussed in the previous section. In order to do this, we include additional white noise terms to the price processes of Eq. (\ref{Itosim}) as described in the previous sub-section (c.f. Eq. (\ref{Logcont})). In this particular example we choose the variance to be $\eta^2 = 10^{-5}$. We then apply the noise to the data of the previous example. In figure 4 we show the result of the estimate of ${\cal A}^2$ for the contaminated data. In this case we assume (correctly) that the dimension of the null space is $k = 2$. We can clearly see how the signal is now negative on average, due to the microstructure noise.  This matches the theoretical prediction in Eq. (\ref{Acont}). In figure 6 we plot the product of subsequent log returns according to (c.f. Eq. (\ref{Logcorr}))
\be  \label{etahat} \hat \eta^2(t) := -\frac{1}{N-1} \sum_{i = 1}^{N-1} \log\left(\tilde X_i(t+1)/\tilde X_i(t)\right)\log\left(\tilde X_i(t)/\tilde X_i(t-1)\right) \;,\ee
where $\tilde X_i$ are the contaminated prices.
According to Eq. (\ref{Logcorr}) of the previous sub-section, we should have $\mathbb{E}[\hat\eta^2] = \eta^2$. This is precisely what we observe in figure 6. In the next subsection, we will see that such signals are very typical of high frequency security prices.

\myfig{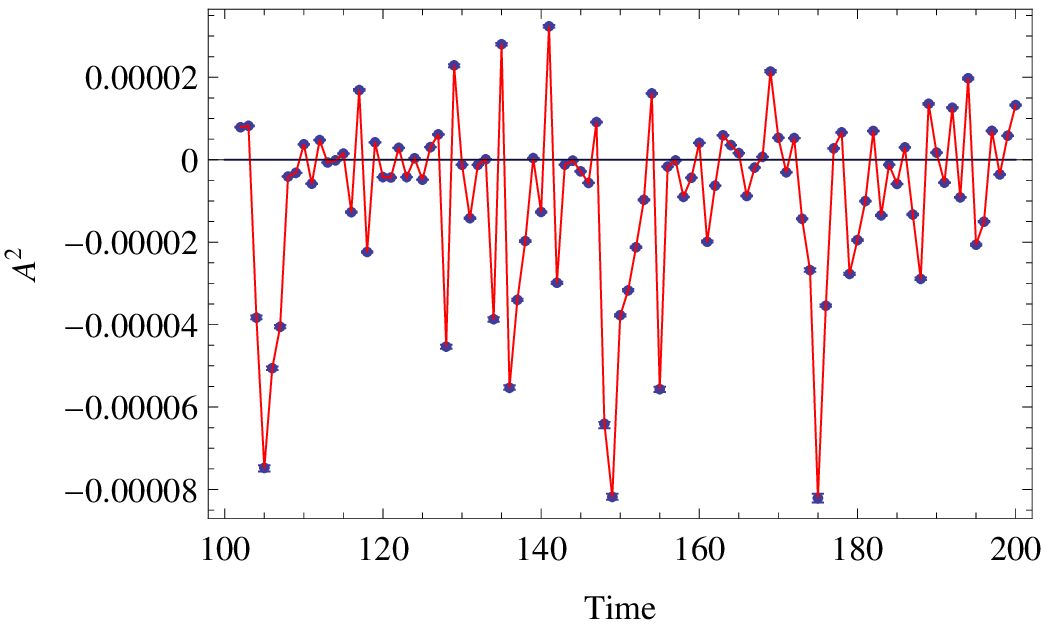}{12}{\small \it Result for the arbitrage
detection algorithm applied to the simulated data in figure 1, now including microstructure noise with variance $\eta^2 = 10^{-5}$. Here we assume (correctly) $k = 2$ null directions.}
\myfig{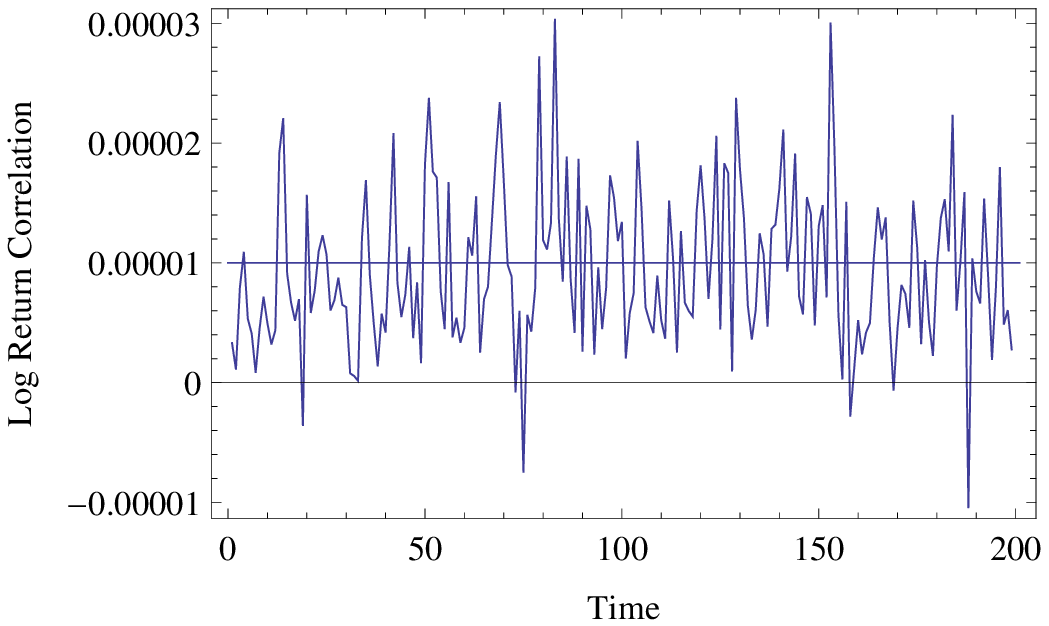}{12}{\small \it Product of subsequent log returns in the  presence of microstructure noise, according to Eq. (\ref{etahat}). The average of this signal is equal to $\eta^2 = 10^{-5}$ (horizontal line) in agreement with the theoretical prediction. }

\subsection{Market Data}
In this section we present some examples of our arbitrage detection
algorithm applied to real financial data. We begin with a look at  three major US
stock indexes: the Dow Jones Composite Average (DJA), the  NASDAQ Composite Index (IXIC) and the NYSE Composite Index  (NYA). Due to their similar nature, we expect strong correlations between these indexes.
Our first sample consist of daily closing prices from September 1
2004 to July 16 2009, a total of 1227 data points. The gauge invariant matrix $\hat G$ has been
estimated using a moving window of 500 days. We have found the
following values for the eigenvalues: $\lambda_1 \approx  2 \times 10^{-3}$,
$\lambda_2 \approx 5\times 10^{-3}$, $\lambda_3 \approx 2\times 10^{-2}$. Therefore, it is reasonable to
assume that the null space has only one dimension, $k = 1$. The
result of the arbitrage detection  algorithm is shown in figure 7. We can see that
the signal is indeed gauge invariant to a very good accuracy. In
figure 7 we have also included an estimate for the gauge invariant
noise term described in section 4.2. In this case we have assumed
that the average of the signal is zero (i.e. no-arbitrage), and so
our estimate for the expected noise is \be \label{noiseest1}
\text{noise}_\pm(t+1) =\pm
\sqrt{\text{Var}_t[\hat{\mathcal{A}}^2(t+1)]} = \pm
\sqrt{\sum_{A=1}^k (\hat\alpha^A(t))^2 \lambda_A(t)}  \;.\ee Looking
at figure 7, we see that the noise can explain most of the signal.
Therefore, we find that our results are consistent with
$\mathcal{A}^2 = 0$, and hence no arbitrage. In figure 8 we show a
histogram of the different values of $\hat{\mathcal{A}}^2$.
As pointed out above, the signal is consistent with $\mathcal{A}^2 =
0$ since the signal to noise ratio is very low:
$\mathbb{E}[\hat{\mathcal{A}}^2]/
\sqrt{\text{Var}[\hat{\mathcal{A}}^2]} \approx -0.0709$.

It is very instructive to look at the trading strategy exploiting the
arbitrage discussed in Proposition \ref{Prop41}.1. In discrete time, the initial value of this portfolio is $V(0) = \sum_\mu
\phi_\mu(0) X_\mu(0) = 0$ and the value at time $t$ is simply
\be \label{alphaint} V(t) =  \sum_{s = 0}^{t-1} \sum_{A,\mu} \hat \alpha^A(s) \hat J_\mu^A(s) \frac{X_\mu(s+1) - X_\mu(s)}{X_\mu(s)} = \sum_{s = 0}^{t-1}  \hat {\cal A}^2(s+1)\;.\ee
In figure 9 we show the value of this portfolio for the daily data of the three US indexes. We include the integrated profit and losses of the indexes themselves for comparison. We have multiplied the index signals by
a numerical factor so that it fits in the same picture. Therefore, the overall scale in the y-axis is irrelevant.   We can see that, as expected, that the performance of this portfolio is very poor for such low frequency data.

\myfig{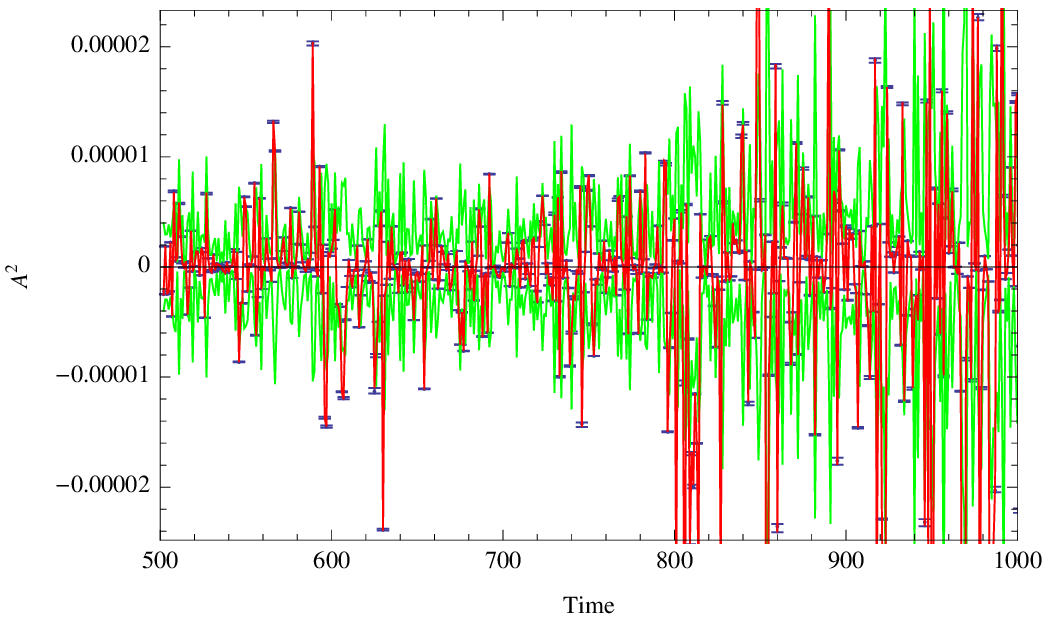}{14}{\it \small Arbitrage detection
algorithm applied to daily closing prices of 3 major US indexes: DJA, IXIC and NYA. Here we show a sample of 500 data
points.  The green lines are an estimate of the variance of the gauge
invariant noise using Eq. (\ref{noiseest1}). The error bars give the
range of values obtained using the different gauges. The solid dots
are the mean of all results. The red line is the USD signal.}

\myfig{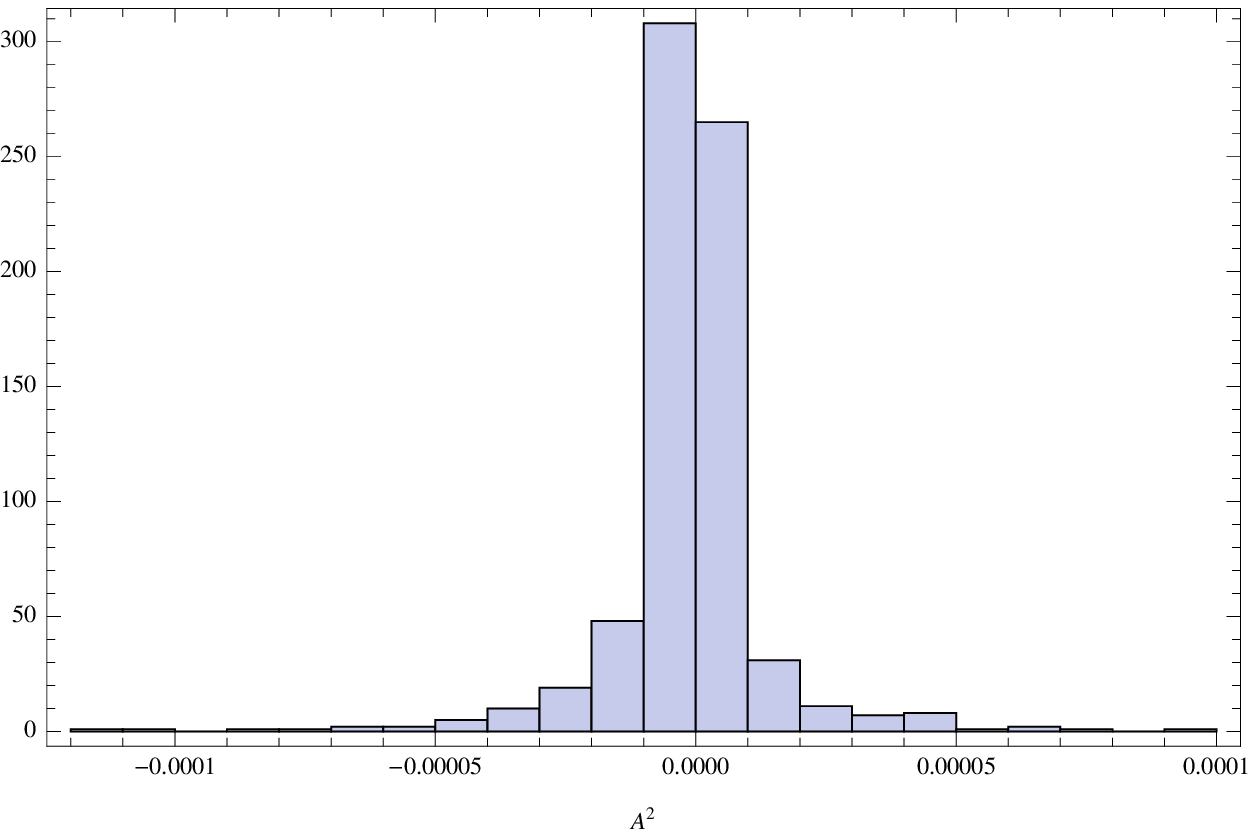}{10}{\it \small Histogram of different
values of $\hat{\cal A}^2$ obtained using daily market data for DJA, IXIC and NYA. The signal to noise ratio is: $\mathbb{E}[\hat{\cal A}^2]/
\sqrt{\text{Var}[\hat{\cal A}^2]} \approx -0.0709$. }

\myfig{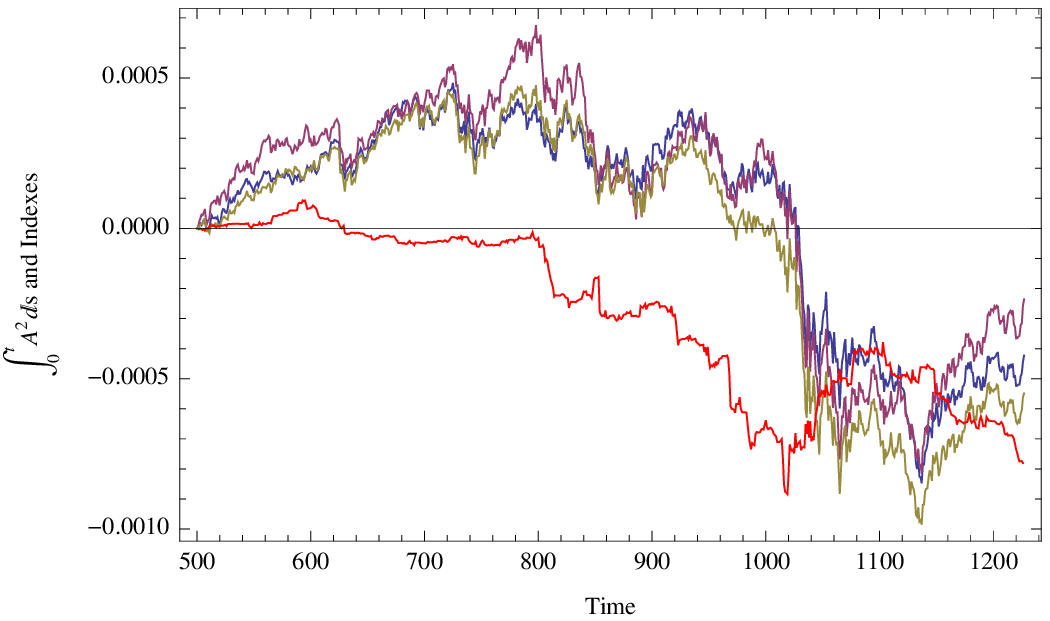}{14}{\it \small  Integrated profit and
losses of the arbitrage portfolio (red line) for the daily data of the US stock indexes  DJA, IXIC and NYA.  We also show the integrated profit and losses of the indexes themselves.  }


Next we look at the same  index set (DJA, IXIC, NYA), but now  at
short time scales. As an example, we study high frequency data obtained on July 28, 2009. The data points have 7 to 10 seconds in separation.
The data was collected using the ``FinancialData" package of
Mathematica. The gauge invariant matrix $\hat G$ has been
estimated using a moving window of 500 data points. We have also assumed one null direction ($k = 1$).  A sample of the arbitrage detection  algorithm is shown in figure 10.  It is quite obvious from this figure that the signal has a very significant positive skewness. In fact, a prominent feature of the signal is a series of positive peaks.  These transient events have a
duration of the order of 5-10 time steps, which for this data is
about one minute. The amplitude of the peaks is quite
significant compared to the noise.  We argue that these peaks are
precisely temporary fluctuations with $\mathcal{A}^2 \neq 0 $, that
is, non-zero curvature events in the market. To show that these are
not isolated events, figure 11 shows the histogram for the full data sample. We can see significant
positive skewness in the signal, compared to the daily data (c.f.
figure 7). In fact, we find a significant signal to noise ratio: $\mathbb{E}[\hat {\cal A}^2]/\sqrt{\text{Var}[\hat {\cal A}^2]} \approx 0.32$. The integrated profit and losses of the arbitrage porfolio of Eq. (\ref{alphaint}) is shown in figure 12. We can see a very good performance in comparison with the daily data (c.f. figure 9). Because of
model risk, such portfolio  can indeed have a
finite probability of a loss on short time scales. However, we see
that on longer scales (integrated signal), the probability of a loss
goes to zero asymptotically as $t\rightarrow \infty$. This is an
example of a {\it statistical arbitrage} as discussed in \cite{Po07,
Bo03}.

We have also studied the effect of the microstructure noise on the high frequency signal. In particular, we have computed the estimate of the noise $\hat \eta^2$ defined in Eq. (\ref{etahat}). We have found that for this particular data sample, the contribution from such noise is very low: $\mathbb{E}[\hat \eta^2]/\sqrt{\text{Var}[\hat \eta^2]} \approx 0.04$. The effect becomes quite significant, however, if we look at traded assets such as stocks and futures.

\myfig{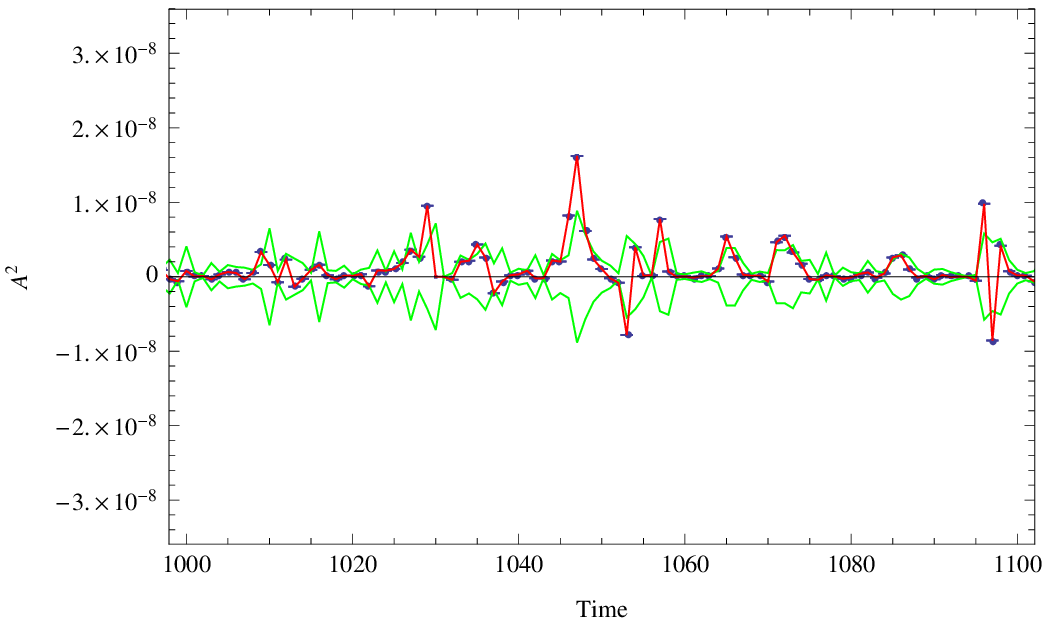}{14}{\it \small Arbitrage detection
algorithm applied to high frequency data for  3 major US indexes: DJA, IXIC and NYA. Here we show a sample of 100 data
points.  The green lines are an estimate of the variance of the gauge
invariant noise using Eq. (\ref{noiseest1}). The error bars give the
range of values obtained using the different gauges. The solid dots
are the mean of all results. The red line is the USD signal.}

\myfig{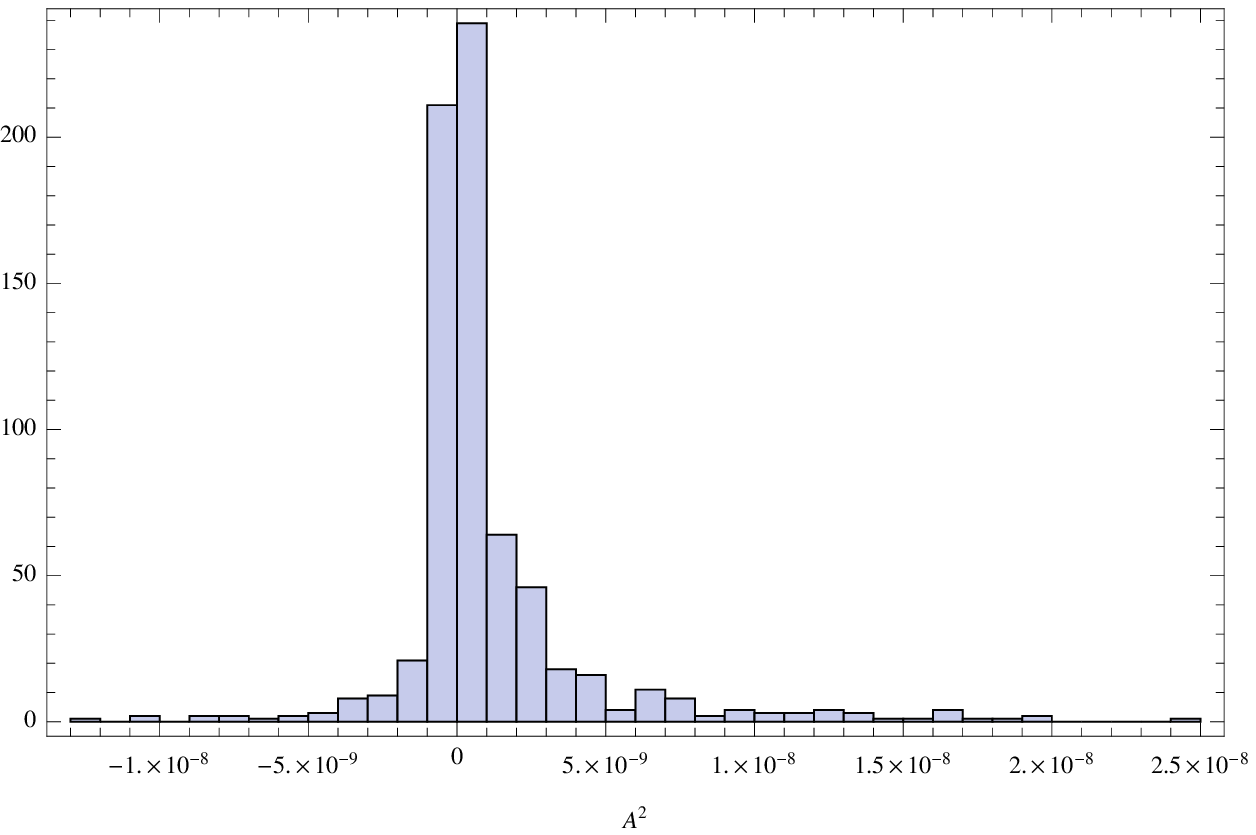}{10}{\it \small Histogram of $\hat{\cal A}^2$ obtained using high frequency data for DJA, IXIC and NYA. The signal to noise ratio is: $\mathbb{E}[\hat{\cal A}^2]/
\sqrt{\text{Var}[\hat{\cal A}^2]} \approx 0.32$. }

\myfig{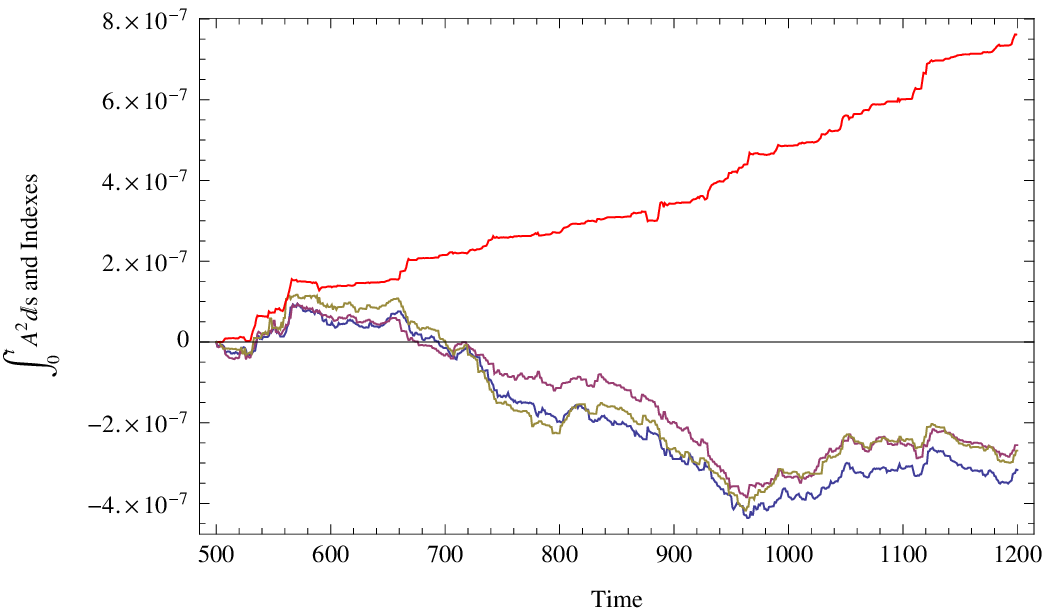}{14}{\it \small Integrated profit and
losses of the arbitrage portfolio (red line) for the high frequency data of the US stock indexes  DJA, IXIC and NYA.  We also show the integrated profit and losses of the indexes themselves. }


Our next data sample consist of the following set of US index futures: E-Mini S\&P 500 (ESU09.CME), DJIA mini-sized (YMU09.CBT), E-Mini Nasdaq 100 (NQU09.CME) and S\&P 500 Index Future (SPU09.CME). The data was collected on August 9, 2009, and all futures expire on September 2009. We have collected prices with a frequency of 7-10 seconds separation, using the ``FinancialData" package of Mathematica.  These securities are highly correlated. Therefore, they are ideal for the search of the arbitrage signal. However, since these are traded instruments, the effect of the bid/ask spread is more pronounced. In figure 13 we show the histogram of the values of $\hat {\cal A}^2$ obtained by applying exactly the same algorithm as in the previous example.  We  get a very poor signal; in fact, we get a negative mean: $\mathbb{E}[\hat{\cal A}^2]/
\sqrt{\text{Var}[\hat{\cal A}^2]} \approx -0.061$. The integrated profit and losses of the simple portfolio of Eq. (\ref{alphaint}) are shown in figure 14.

We have also calculated the effect of the microstructure noise, by computing the estimate $\hat \eta^2$ defined in Eq. (\ref{etahat}).  The effect for this data sample is about an order of magnitude bigger than the previous example: $\mathbb{E}[\hat \eta^2]/\sqrt{\text{Var}[\hat \eta^2]} \approx 0.15$. This noise is the main obstacle to a detection of ${\cal A}^2$. Nevertheless, one can devise more complicated detection methods which filter out the microstructure noise. In figure 14 we show the integrated profit and losses of a particular proprietary strategy. A detailed study of such strategies is beyond the scope of this paper.

\myfig{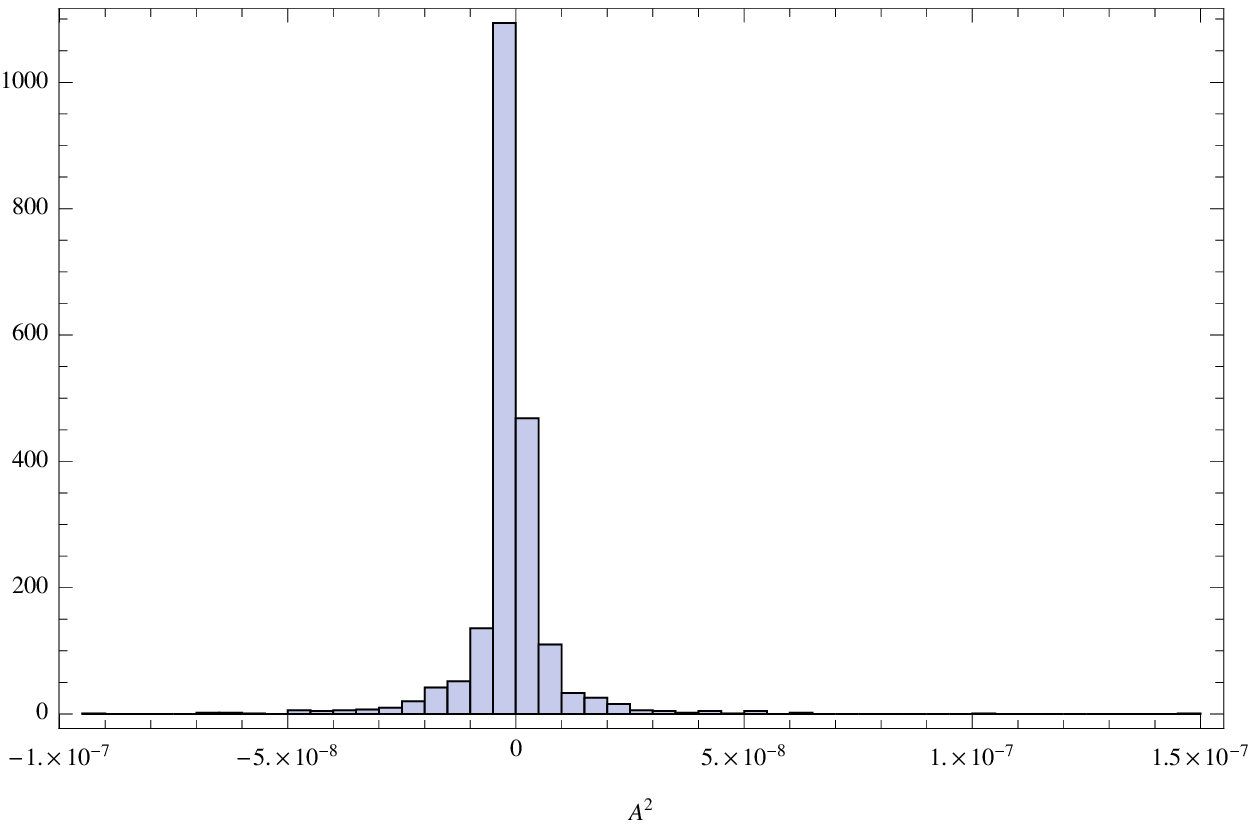}{10}{\it \small Histogram of $\hat{\cal A}^2$ obtained using high frequency data for the index futures: ESU09.CME, YMU09.CBT, NQU09.CME, SPU09.CME. The signal to noise ratio is: $\mathbb{E}[\hat{\cal A}^2]/
\sqrt{\text{Var}[\hat{\cal A}^2]} \approx -0.061$. This picture illustrates the negative effects of the market microstructure noise for the simplest trading strategy.}

\myfig{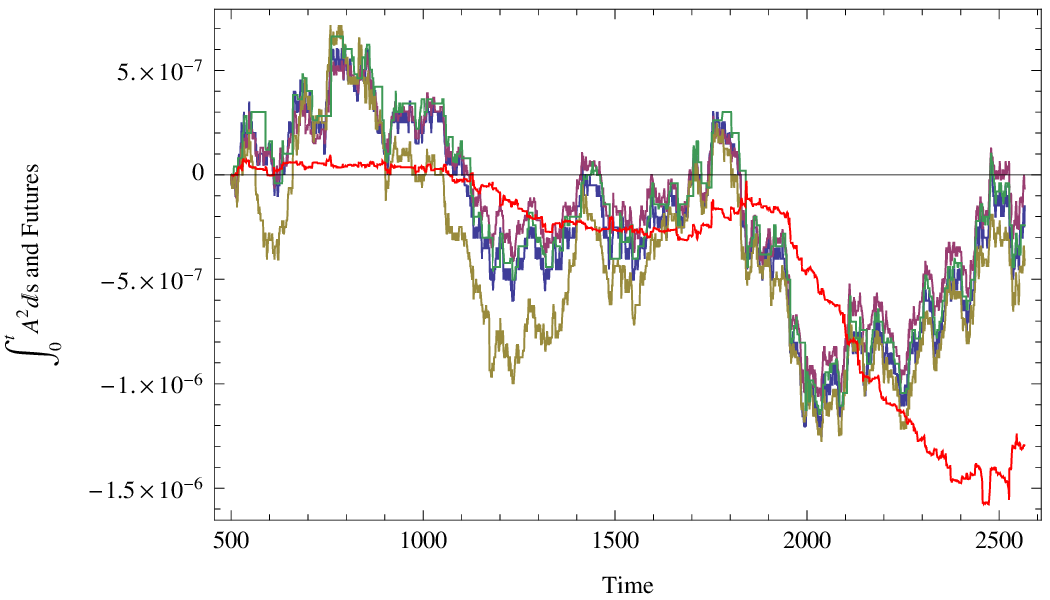}{14}{\it \small Integrated profit and
losses of the arbitrage portfolio (red line) of Eq. (\ref{alphaint}) for the high frequency data of the US index futures: ESU09.CME, YMU09.CBT, NQU09.CME, SPU09.CME.  We also show the integrated profit and losses of the futures themselves. This picture illustrates the negative effects of the market microstructure noise for the simplest trading strategy. }

\myfig{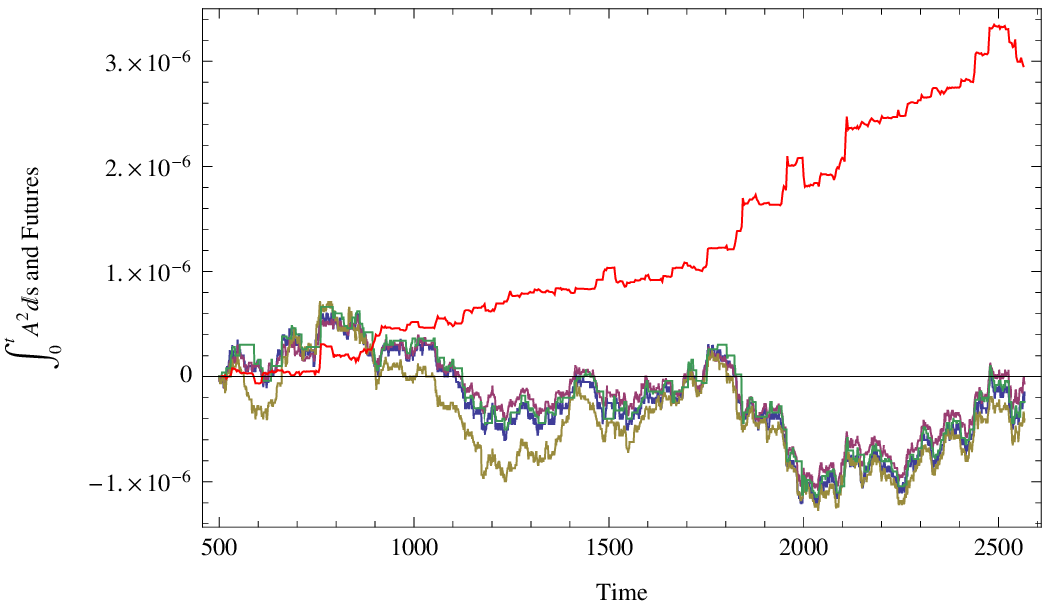}{14}{\it \small Integrated profit and
losses of a proprietary arbitrage portfolio (red line) for the high frequency data of the US index futures: ESU09.CME, YMU09.CBT, NQU09.CME, SPU09.CME.  We also show the integrated profit and losses of the futures themselves.  This particular strategy is designed to minimize the effects of the microstructure noise.}

\section{Conclusions}

In this paper we have defined a general measure of arbitrage which
is invariant under changes of num\'{e}raire and equivalent
probability measure. Our main assumption is that all financial
instruments can be described by It\^o processes. This is not a very
strong assumption as many complex financial models, including those
reflecting the non-Gaussian nature of stock returns, can be modeled
 this way. We showed that the gauge invariant arbitrage measure can
be interpreted in terms of the curvature of the stochastic version
of the Malaney-Weinstein connection (\cite{Ma96}, \cite{We06}). The
zero curvature condition is then equivalent to the no-arbitrage
principle. Moreover, we demonstrated a simple generalization of the
classic asset pricing theorem to include arbitrage. Finally, we have
presented a basic algorithm to measure the market curvature using
financial data. We found evidence for non-zero curvature
fluctuations in high frequency data involving stock indexes and index futures.\par From a financial perspective, our algorithms can
be used to exploit arbitrage systematically, and generate profitable
trading strategies. This will require much more empirical research,
and the development of more sophisticated techniques to estimate the
arbitrage curvature measure $\mathcal{A}^2$. This is left for future
work.\par From a scientific perspective, we believe that our
findings represent a modest step towards the understanding of the
non-equilibrium market dynamics. Gauge theories provide the natural
mathematical language to that aim, and arbitrage opportunities can
be interpreted as a non-zero curvature fluctuation in an economy out
of equilibrium. It is interesting how most of our current economic
and financial thinking rely so much on the assumptions of general
equilibrium theory.\par There has been a growing consensus that we
need a better understanding of the non-equilibrium dynamics of the
economy (see e.g. \cite{FaGe08}). In particular, one would like to
understand what is the relaxation time scale for non-equilibrium
fluctuations to disappear (if they do). Within the limited data
sample that we have shown in this paper, the relaxation time seems
to be of the order of one minute.  However, this can be very
different in other sectors of the market.

\section*{Acknowledgements}
We would like to extend our gratitude to Eric Weinstein, Pia Malaney, Lee Smolin, Bernd Schroers, Mike Brown, Simone Severini, Jiri Hoogland,  Jim Herriot, and Bruce Sawhill for many discussions and  ideas which influenced the results of this paper. Research at
Perimeter Institute is supported by the Government of Canada through
Industry Canada and by the Province of Ontario through the Ministry
of Research \& Innovation.

\section*{Appendix A}
In this appendix we consider the following general mathematical
problem. Suppose we are given two $N\times N$ symmetric and real
matrices which we label by $G(t)$ and $G(t+1)$ respectively.
Moreover, both matrices have a null space with the same
dimensionality. In other words, let $J^A(t)$ and $J^A(t+1)$ be
orthonormal basis vectors such that \be G(t) \ J^A(t) = 0\;,\;\;\;
A=1,\ldots,k\;,\;\;\; G(t+1) \ J^B(t+1) = 0\;,\;\;\; B = 1,\ldots,
k'\;,\ee and $k = k'$. Now let $|| \cdot ||$ be some matrix norm
such that $|| G(t) - G(t+1)|| \ll 1$. More precisely, we are
interested in the limit in which both matrices are very similar to
each other. We then wish to find a $k\times k$ rotation matrix $C$
such that \be \lim_{|| G(t) - G(t+1)|| \rightarrow 0} J^A(t+1)=
\sum_B C^{A B} J^B(t)\;,\;\;\; C^\dagger\ C = \mathbf{1}\;.\ee In
physics, this problem arises in the so-called {\it degenerate
perturbation theory} \cite{Sa94}.  However, it is often assumed that
$G(t+1)$ will ``lift the degeneracy", which in our case means $k' <
k$. This is not the case in our problem, and so the usual
perturbation theory methods do not work.

We are interested in solving the problem above only approximately.
In fact, in practice we have no control over $G(t)$ and $G(t+1)$,
and so we cannot take the limit  $|| G(t) - G(t+1)|| \rightarrow 0$
explicitly. However, we will {\it assume} that $|| G(t) -
G(t+1)||\ll 1$ (which is true most of the time). Now consider the $k
\times k$ real matrix $\tilde C$ with components \be \tilde C^{A B}
:= \left[ J^A(t+1)\right]^\dagger \  J^B(t)\;.\ee Under our
assumptions, we have \be \tilde C^{A B} \approx C^{A B} + {\cal
O}(\delta G)\;,\ee where $\delta G = G(t) - G(t+1)$. Therefore, we
seek a rotation matrix $\hat C$ which approximates $\tilde C$ as
closely as possible. In the limit $|| G(t) - G(t+1)||\rightarrow 0$
we should have $\hat C \rightarrow C$. In other words, $\hat C$ will
be the our best approximation to $C$.

We can solve this problem using Lagrange multipliers. We seek to
minimize the Lagrangian \be {\cal L} = \tr \left[ (\tilde C-\hat
C)^\dagger \ (\tilde C-\hat C)\right] +\tr\left[\lambda \ (\hat C^T\
\hat C - \mathbf{1}) \right]\;,\ee where $\lambda$ is a symmetric
matrix which serves as a Lagrange multiplier implementing the
constraint $\hat C^T \ \hat C= \mathbf{1}$. The Lagrange equations
are then \be \label{Leq} \frac{\partial {\cal L}}{\partial \hat C^{A
B}} = 0\;,\;\;\;  \frac{\partial {\cal L}}{\partial \lambda^{A B}} =
0\;.\ee We will only present the solution to Eqs. (\ref{Leq}) for
the cases of $k = 1$ and $k = 2$.

For $k = 1$, the rotation matrix $\hat C$ is simply a sign and one can show that the minimum of ${\cal L}$ is
\be \hat C =  \text{sign}(\tilde C)\;.\ee
For $k = 2$, the solution can be written in terms of the Pauli matrices:
\be \sigma_0 :=
 \begin{pmatrix}
1 & 0 \\
0 &1
\end{pmatrix} \;,\;\;\; \sigma_1 :=
 \begin{pmatrix}
0 & 1 \\
1 &0
\end{pmatrix}\;,\;\;\;
\sigma_2 :=
 \begin{pmatrix}
1 & 0 \\
0 &-1
\end{pmatrix}\;,\;\;\;
\sigma_3 :=
 \begin{pmatrix}
0 & 1 \\
-1 &0
\end{pmatrix}\;.
\ee We can expand, \be \tilde C = \sum_{i = 0}^3 \tilde c_i
\sigma_i\;, \;\;\; \hat C = \sum_{i = 0}^3 \hat c_i
\sigma_i\;,\;\;\; \lambda = \sum_{i = 0}^2 \lambda_i \sigma_i\;,\ee
where \be \tilde c_i = \frac{1}{2}\tr(\sigma_i^\dagger \ \tilde
C)\;,\;\;\; i=0,\ldots, 3\;,\ee and similarly for $\hat c_i$ and
$\lambda_i$. The solution to the Lagrange equations, Eq.
(\ref{Leq}), which correspond to the minimum of ${\cal L}$ can be
shown to be \be \hat c_0 = \hat c_3 = 0\;,\;\;\; \hat c_1 =
\frac{\tilde c_1}{\sqrt{\tilde c_1^2 + \tilde c_2^2}}\;,\;\;\; \hat
c_2 = \frac{\tilde c_2}{\sqrt{\tilde c_1^2 + \tilde c_2^2}}\;,\ee
for $\tilde c_1^2 + \tilde c_2^2 > \tilde c_0^2 + \tilde c_3^2$, and
\be \hat c_1 = \hat c_2 = 0\;,\;\;\; \hat c_0 = \frac{\tilde
c_0}{\sqrt{\tilde c_0^2 + \tilde c_3^2}}\;,\;\;\; \hat c_3 =
\frac{\tilde c_3}{\sqrt{\tilde c_0^2 + \tilde c_3^2}}\;,\ee
otherwise.

\end{document}